\documentclass[]{JHEP3} 

\usepackage{epsfig,multicol,bbm,multirow}
\usepackage[sort&compress,numbers]{natbib}
\usepackage{subfigure}
\usepackage{graphicx}

\newcommand\fverb{\setbox\fverbbox=\hbox\bgroup\verb}
\newcommand\fverbdo{\egroup\medskip\noindent%
			\fbox{\unhbox\fverbbox}\ }
\newcommand\fverbit{\egroup\item[\fbox{\unhbox\fverbbox}]}
\newbox\fverbbox

\usepackage{graphicx}
\def\beq{\begin{equation}}
\def\eeq{\end{equation}}

\def\prop#1{{\cal P}_{#1}}

\def\Ls{\mathrm{Ls}}

\def\I33m{\mathrm{I}_3^{3{\mathrm m}}}

\def\fl{{\rm f}}
\def\be{\begin{equation}}
\def\ee{\end{equation}}
\def\bea{\begin{eqnarray}}
\def\eea{\end{eqnarray}}

\def\qb{{\bar{q}}}
\def\cg{c_\Gamma}

\def\eps{\epsilon}

\def\musq{\mu^2}
\def\spa#1.#2{\left\langle#1#2\right\rangle}
\def\spb#1.#2{\left[#1#2\right]}
\def\lor#1.#2{\left(#1#2\right)}
\def\sand#1.#2.#3{%
\left\langle\smash{#1}{\vphantom1}^{-}\right|{#2}%
\left|\smash{#3}{\vphantom1}^{-}\right\rangle}
\def\sandp#1.#2.#3{%
\left\langle\smash{#1}{\vphantom1}^{-}\right|{#2}%
\left|\smash{#3}{\vphantom1}^{+}\right\rangle}
\def\sandpp#1.#2.#3{%
\left\langle\smash{#1}{\vphantom1}^{+}\right|{#2}%
\left|\smash{#3}{\vphantom1}^{+}\right\rangle}
\def\sandpm#1.#2.#3{%
\left\langle\smash{#1}{\vphantom1}^{+}\right|{#2}%
\left|\smash{#3}{\vphantom1}^{-}\right\rangle}
\def\sandmp#1.#2.#3{%
\left\langle\smash{#1}{\vphantom1}^{-}\right|{#2}%
\left|\smash{#3}{\vphantom1}^{+}\right\rangle}
\def\spab#1.#2.#3{\langle#1|#2|#3]}
\def\spba#1.#2.#3{[#1|#2|#3\rangle}
\def\spaa#1.#2.#3{\langle#1|#2|#3\rangle}
\def\spbb#1.#2.#3{[#1|#2|#3]}
\def\spaxa#1.#2.#3.#4{\langle#1|#2|#3|#4\rangle}
\def\spbxb#1.#2.#3.#4{[#1|#2|#3|#4]}

\def\fig{Fig.}

\def\aa{\mathcal{A}}
\def\lzero#1{\mathrm{L}_0 \left( #1 \right)}
\def\lone#1{\mathrm{L}_1 \left( #1 \right)}
\def\lc{\mathrm{lc}}
\def\slc{\mathrm{sl}}
\def\fl{\mathrm{fl}}
\def\tree{\mathrm{tree}}
\def\nlf{n_{\mathrm{f}}}
\def\vv{\mathrm{v}}
\def\ax{\mathrm{ax}}

\title{Next-to-leading order predictions for $Z \gamma+$jet and $Z \gamma \gamma$
final states at the LHC.}

\author{
    John M. Campbell$^{a}$, Heribertus B. Hartanto$^{a,b}$ and Ciaran Williams$^{a}$
    \\
    $^{a}$Fermilab, Batavia, IL 60510, USA \\
    $^{b}$Physics Department, Florida State University,  Tallahassee, FL 32306, USA	   	
    \\
    E-mails: 
    {\tt johnmc@fnal.gov}, 
    {\tt hhartanto@hep.fsu.edu}, 
    {\tt ciaran@fnal.gov}.}

\received{\today} 		
\accepted{\today}		

\preprint{
FERMILAB-PUB-12-408-T \\
FSU-HEP-120802 \\
}

\abstract{We present next-to-leading order predictions for final states containing leptons produced through the decay of a $Z$
boson in association with either a photon and a jet, or a pair of photons. The effect of photon radiation from the final
state leptons is included and we also allow for contributions arising from fragmentation processes.  Phenomenological studies are presented for
the LHC in the case of final states containing charged leptons and in the case of neutrinos. We also use
the procedure introduced by Stewart and Tackmann to provide a reliable estimate of the scale uncertainty inherent in
our theoretical calculations of jet-binned $Z\gamma$ cross sections.
These computations have been implemented in the public code MCFM.}


\begin{document} 


\section{Introduction}
The production of vector bosons at hadron colliders provides a stringent testing ground for the Standard Model. The
presence of a $W$ or $Z$ boson ensures that such processes are sufficiently hard to be reliably
computed in perturbative QCD. The fact that the final states are produced via weak couplings means that it is a
tough experimental challenge to observe these processes on top of the much larger QCD-dominated backgrounds. The
search for processes involving multiple vector bosons also provides a crucial test of a key element of the Standard
Model, namely interactions between the vector bosons themselves. Measuring these cross sections also allows one to
place limits on additional self-couplings that may be induced by effective operators in many extensions of the
Standard Model.

Theoretical predictions for the hadronic production of a $Z$ boson and a photon have been significantly improved beyond the tree-level
approximation. Next-to-leading order (NLO) QCD corrections~\cite{Ohnemus:1992jn,Baur:1997kz,Campbell:1999ah,DeFlorian:2000sg,Campbell:2011bn} have been
supplemented by higher-order gluon-initiated contributions~\cite{Ametller:1985di,vanderBij:1988fb,Adamson:2002rm,Campbell:2011bn} and NLO
electroweak effects have also been included~\cite{Hollik:2004tm,Accomando:2005ra}. These predictions, for the inclusive production of a
$Z\gamma$ final state, have been confronted with data from both the Tevatron~\cite{Aaltonen:2011zc,Abazov:2011qp} and the
LHC~\cite{Chatrchyan:2011rr,Aad:2011tc,Aad:2012mr} experiments. However, particularly in the LHC environment, it can be advantageous to
perform a less inclusive measurement in order to more cleanly isolate a potential signal. In particular it has become a common practice
to separate an analysis into categories that are classified by the number of jets identified in the final state. Such an analysis is
frequently referred to as a binned-analysis. This type of analysis can, for instance, be optimized to take advantage of the different
expected background contributions in the various bins. Whilst this is a useful experimental tool it introduces new theoretical
difficulties. This is due to the fact that the binning procedure introduces a new kinematic scale, namely the transverse momentum cut used
to define a jet, upon which the theoretical prediction must depend. One normally expects this richer kinematic structure to lead to a less
reliable perturbative calculation.

At leading order (LO) the theoretical prediction for the $Z\gamma$ cross section contains final states including only the $Z$ and the
photon. At NLO one includes virtual corrections to the LO topology and bremsstrahlung events corresponding to the emission
of an additional parton. When the measurement is performed using exclusive jet-bins all of the NLO aspects of the
calculation are confined to the 0-jet bin. Specifically  the 0-jet bin includes the virtual corrections and the real
emissions for which the parton is unresolved. The theory prediction for the remaining 1-jet bin is identical to
performing a LO $Z\gamma$+jet calculation and ignoring the 0-jet bin altogether.  In summary, the NLO calculation of the
inclusive $Z\gamma$ cross section provides a NLO prediction for the 0-jet bin, a LO prediction for the 1-jet bin
and the prediction for bins with jet multiplicity of two or more is zero. 

The most natural way to improve the theory is to incorporate higher order perturbative corrections into the inclusive
cross section prediction. This is a difficult task, although progress on this front may be expected in the relatively near future
(see for example Ref.~\cite{Catani:2011qz} for NNLO predictions for the diphoton process). In lieu of such a
calculation one can instead focus on improving the theoretical predictions in the 1-jet bin. By performing a separate
NLO calculation for the $Z\gamma$+jet final state, the 1-jet bin can be predicted at the NLO level and a non-zero prediction
is obtained in the 2-jet bin. This covers the kinematic range of the NNLO prediction for the $Z\gamma$ rate and has the same
accuracy in the 1- and 2-jet bins. Such a calculation does not, of course, include any higher order effects in the 0-jet bin 
and therefore the accuracy of the $Z\gamma$ inclusive cross section is not improved.

To this end, in this paper we present NLO corrections to the final state consisting of a $Z$ boson, a photon and a jet.
Although we will often refer to it in these terms (``$Z\gamma$+jet''), in fact we will actually compute the NLO
corrections to the process,
\begin{equation}
p + p \rightarrow  \ell\bar{\ell}+ \gamma + \mbox{jet} \;,
\label{procZgamjet}
\end{equation}
where the leptons are produced from either a $Z$ boson or a virtual photon $\gamma^*$. We will consider both charged
and neutral leptonic decays  i.e. $\ell = e, \nu$.  When $\ell$ is a charged lepton we include the contributions in
which the photon is radiated  from the leptons. It is for this reason that the terminology ``$Z\gamma$+jet'' 
is misleading, since it suggests the  production of a $Z$ boson, a jet and a photon, with the subsequent decay of the
$Z$ factorized from the process.  As in the most recent study of the $Z\gamma$ process~\cite{Campbell:2011bn}, we
include the effects of photon fragmentation in order to allow for photon isolation criteria that are currently used in
experimental studies.

As a by-product of this calculation, we shall also present results for the ``$Z\gamma\gamma$'' process,
\begin{equation}
p + p \rightarrow  \ell \bar{\ell} + \gamma + \gamma \;,
\label{procZgamgam}
\end{equation}
where either, or both, of the photons may be radiated from a charged lepton.
Many of the amplitudes relevant for this calculation can be obtained from the $Z\gamma$+jet case
by extracting the subleading-in-colour contributions. Although this process has previously been computed at
NLO~\cite{Bozzi:2011en}, we extend that treatment slightly by including  fragmentation contributions.

Our paper is structured as follows. In sections~\ref{sec:zajcalc} and~\ref{sec:zaacalc} we describe the analytic calculation
of the  $Z\gamma$+jet and $Z\gamma\gamma$ processes respectively. Section~\ref{sec:photiso} discusses the issues of
photon isolation and fragmentation. 
We present our results and some phenomenological examples at the LHC in Section~\ref{sec:zaj_ph} and
summarize our findings in Section~\ref{sec:conc}.
Finally in Appendices~\ref{app:tree} and~\ref{app:zgamgam} we present formulae for the helicity amplitudes that are used in our calculations.

\section{Calculation of $Z \gamma+$jet amplitudes}
\label{sec:zajcalc}

In this section we present details of the analytic calculation of the NLO corrections to the $Z\gamma$+jet process. Although 
NLO results exist in the literature for the $W\gamma$+jet and $W\gamma\gamma$+jet final states~\cite{Campanario:2009um,Campanario:2011ud},
the calculation of the $Z\gamma$+jet process considered here is new. We include more detailed results in Appendix~\ref{app:tree}. 
Note that we do not include higher-order finite contributions of the form, $gg \to Z\gamma g$, that contribute at the level
of a few percent at the LHC~\cite{Agrawal:2012df}. 

\begin{figure}
\begin{center}
\includegraphics[width=14cm]{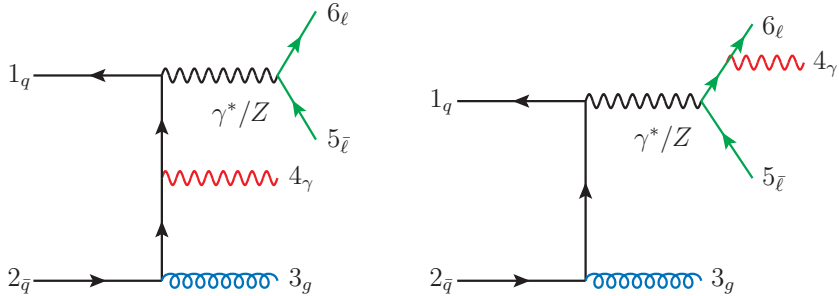}
\caption{Examples of leading order diagrams for the $Z\gamma$+jet process, for the cases of
photon emission from the quark line (``$q$-type'', left) and from the lepton line (``$\ell$-type'', right).
The particle labels correspond to momentum assignments that are all outgoing. }
\label{fig:zajlo}
\end{center}
\end{figure}
Tree-level topologies associated with the hadronic production of a $Z$ boson (with subsequent leptonic decays),
a photon and a jet are shown in Fig.~\ref{fig:zajlo}. As illustrated in the figure, there are 
two separately gauge invariant classes of Feynman diagrams in the $Z\gamma$+jet process. These can be classified
according to whether the photon is:
\begin{itemize}
\item emitted from the quark line, ``$q$-type'' (Fig.~\ref{fig:zajlo}, left),
\item emitted from the lepton line, ``$\ell$-type'' (Fig.~\ref{fig:zajlo}, right).
\end{itemize}
Clearly when the $Z$ decays to neutrinos only the $q$-type diagrams are present. 
Using the above nomenclature we can write the tree-level $0 \rightarrow q \qb g \gamma \bar{\ell} \ell$ amplitude  
in a form in which the explicit colour and electroweak charge structures are separated from the kinematics, 
\begin{eqnarray}
&& A^{(0)}(1_q, 2_{\qb}, 3_{g}, 4_{\gamma}, 5_{\bar{\ell}}, 6_{\ell}) = 2 \sqrt{2} e^3 g_s T^{a_3}_{i_1 i_2} \nonumber \\*
&& \times \bigg[ Q_q \left(-Q_q + v_{L,R}^{q}v_{L,R}^{\ell} \prop{Z}(s_{56}) \right) 
  \aa^{(0)}_{q}(1_q, 2_{\qb}, 3_{g}, 4_{\gamma}, 5_{\bar{\ell}}, 6_{\ell}) \nonumber \\*
&& \;\;  +Q_\ell\left(-Q_q + v_{L,R}^{q}v_{L,R}^{\ell} \prop{Z}(t_{456}) \right)
  \aa^{(0)}_{\ell}(1_q, 2_{\qb}, 3_{g}, 4_{\gamma}, 5_{\bar{\ell}}, 6_{\ell}) \bigg] \;.
\label{eq:zajcolor1}
\end{eqnarray}
Here the subscripts $L$ and $R$ refer to the handedness of the fermion that couples to the $Z$.
The QED and QCD couplings are represented by $e$ and $g_s$ respectively and $Q_i$ is the electric charge of particle $i$ in units of $e$.  
The fermionic (quark/lepton) coupling to the $Z$ boson, $v^{(q,\ell)}_h$ is given by
\begin{eqnarray}
  v^{\ell}_L &=& \frac{-1-2Q_{\ell} \sin^2 \theta_W}{\sin 2 \theta_W},   
\hspace{1.7cm} v^{\ell}_R = -\frac{2 Q_{\ell}\sin^2 \theta_W}{\sin 2 \theta_W},  \nonumber\\
   v^q_L &=& \frac{\pm 1 - 2 Q_q \sin^2 \theta_W}{\sin 2 \theta_W},   
\hspace{1.7cm} v^q_R = -\frac{2 Q_q \sin^2 \theta_W}{\sin 2 \theta_W}, 
\end{eqnarray}
where $\theta_W$ is the Weinberg angle. 
The sign in $v^q_L$ distinguishes between up $(+)$  and down $(-)$ type quarks. 
The propagator factor, which is the ratio of the $Z$ and photon propagators, is given by,
\begin{equation}
\prop{Z}(s) = \frac{s}{s-M_Z^2 + i \Gamma_Z M_Z} \;,
\end{equation}
where $M_Z$ and $\Gamma_Z$ are the mass and the width of the $Z$ boson.
We shall present expressions for the helicity amplitudes in standard
spinor notation, with the spinor products defined as, 
\begin{equation}
\spa i.j=\bar{u}_-(p_i) u_+(p_j), \;\;\;
\spb i.j=\bar{u}_+(p_i) u_-(p_j), \;\;\;
\spa i.j \spb j.i = 2 p_i \cdot p_j\,.
\end{equation}
A description of spinor helicity methods can be found in, for instance, Ref.~\cite{Dixon:1996wi}.  
The helicity amplitudes for the $q$-type diagrams can be obtained from the primitive amplitudes 
for the $e^+ e^- \rightarrow q \qb g g$ process presented in Ref.~\cite{bdk1}, once they are
dressed with appropriately-changed color factors. At tree level this amounts to simply symmetrizing over the two gluon
orderings in the partial amplitudes.
For example, the tree-level amplitude $\aa^{(0)}_q(1^{+}_{q},2^{-}_{\bar{q}},3^{+}_{\gamma},4^+_g,5^{-}_{\bar{\ell}},6^+_{\ell})$ 
can be obtained from Eq.~(8.4) of Ref.~\cite{bdk1} that reads,
\begin{eqnarray}
A_6^{\tree}(1^+_q,2^-_{\qb},3^+_g,4^+_g)=-i\frac{\spa2.5^2}{\spa1.4\spa2.3\spa3.4\spa5.6} \;.
\label{eq:bdk_h1} 
\end{eqnarray}
Note that, following the original notation, we have suppressed the explicit dependence on the leptons
in the amplitude definition on the left-hand side. 
Performing the symmetrization one finds,
\begin{eqnarray} 
\aa^{(0)}_q(1^{+}_{q},2^{-}_{\bar{q}},3^{+}_{g},4^+_{\gamma},5^{-}_{\bar{\ell}},6^+_{\ell})
&=&A_6^{\tree}(1^+_q,2^-_{\qb},3^+_g,4^+_g) + A_6^{\tree}(1^+_q,2^-_{\qb},4^+_g,3^+_g)  \nonumber\\
&=&-i\frac{\spa1.2\spa2.5^2}{\spa 1.3\spa1.4\spa2.3\spa2.4\spa5.6} \;,
\label{eq:phot_h1}
\end{eqnarray}
where the simplification results from use of the Schouten identity.
Inspection of the above formula clearly confirms the correct QED structure. 
Compared to the original formula in Eq.~(\ref{eq:bdk_h1}), the QCD pole 
associated with the triple gluon vertex $\spa 3.4$ has disappeared whilst new poles 
$\spa 1.3$ and $\spa 2.4$ have appeared. The new poles are associated 
with the fact that the photon is not colour-ordered. 
The remaining independent tree-level helicity amplitudes can be calculated using exactly the same prescription. 
The resulting expressions, together with the full details of the virtual and real radiation amplitudes, are given
in Appendix~\ref{app:tree}. 

The $\ell$-type amplitudes cannot be extracted directly from the sub-leading colour pieces of the QCD amplitudes. 
This is obvious since the gluon can never be radiated from the leptons. 
However, as we shall explain in more detail below, we can take advantage of the fact that the QCD elements of the amplitude are all 
incorporated in the $0 \to q\bar{q} g \bar{\ell} \ell$ amplitude, and that the electroweak information 
(i.e. $Z\rightarrow \ell^+\ell^-\gamma$) factorizes from the QCD part. 
Therefore we can use the $e^+e^- \rightarrow q \bar{q} g$ amplitudes, given in the same notation in Ref.~\cite{bdk1},
and factor out the $Z\rightarrow \ell^+\ell^-$ current, 
\begin{equation} 
\aa_q(1_q,2_{\bar{q}},3_g,5_{\bar{\ell}},6_{\ell}) = \tilde\aa_{\mu}(1_q,2_{\bar{q}},3_g) J^{\mu}_{\ell\bar{\ell}} (5_{\bar{\ell}},6_{\ell}). 
\label{eq:demo1}
\end{equation}
This extraction of the current from the amplitude is fairly straightforward.
For example, consider the tree level helicity amplitude,
\begin{eqnarray}
\aa_q^{(0)} (1^+_q,2^-_{\bar{q}},3^+_g,5^-_{\bar{\ell}},6^+_{\ell})=-i \frac{\spa 2.5^2}{\spa 1.3\spa2.3\spa5.6} \;.
\label{eq:demo4}
\end{eqnarray}
In order to manipulate this into the desired form we first multiply both the numerator and denominator by $\spb6.5$, in order to
obtain a factor of $s_{56}$ in the denominator that can be factored into the current $J^{\mu}_{\ell\bar{\ell}}$. In the numerator,
we use momentum conservation to write, $\spa2.5 \spb5.6 = -\spab2.{(1+3)}.6$ and obtain,
\begin{eqnarray}
\aa^{(0)}_q (1^+_q,2^-_{\bar{q}},3^+_g,5^-_{\bar{\ell}},6^+_{\ell})&=&
 i \frac{\spaxa2.(1+3).\gamma_{\mu}.2}{2 \spa1.3 \spa2.3} \times \left( - \frac{\spab5.\gamma^{\mu}.6}{s_{56}} \right) \;,
\label{eq:demo5}
\end{eqnarray}
where we have explicitly undone the Fierz identity.
At this point we have identified the current,
\begin{eqnarray} 
J^{\mu}_{\ell\bar{\ell}} (5^-_{\bar{\ell}},6^+_{\ell})&=&  - \frac{\spab5.\gamma^{\mu}.6}{s_{56}}, 
\label{eq:demo3a}
\end{eqnarray}
and the factorization in Eq.~(\ref{eq:demo1}) is manifest with the identification,
\begin{eqnarray}
\tilde\aa^{(0)}_{\mu}(1^+_q,2^-_{\qb},3^+_g) = i \frac{\spaxa2.(1+3).\gamma_{\mu}.2}{2 \spa1.3 \spa2.3} \;.
\end{eqnarray}
The $\ell$-type amplitudes can then be obtained by contracting the $\tilde\aa_{\mu}(1_q,2_{\bar{q}},3_g)$ piece 
with the $Z\rightarrow \ell^+\ell^-\gamma$ current,
\begin{equation}
\aa_{\ell}(1_q,2_{\bar{q}},3_g,4_{\gamma},5_{\bar{\ell}},6_{\ell}) 
= \tilde\aa_{\mu}(1_q,2_{\bar{q}},3_g) J^{\mu}_{\ell\bar{\ell}\gamma} (4_{\gamma},5_{\bar{\ell}},6_{\ell}) \;,
\label{eq:demo2}
\end{equation}
where the currents for the $Z$ boson decay including photon radiation from the leptons are,
\begin{eqnarray} 
J^{\mu}_{\ell\bar{\ell}\gamma} (4^+_{\gamma},5^-_{\bar{\ell}},6^+_{\ell}) &=& 
\frac{\spaxa5.\gamma^{\mu}.(4+6).5}{t_{456} \spa4.5 \spa4.6} \;, 
\label{eq:demo3b} \\
J^{\mu}_{\ell\bar{\ell}\gamma} (4^-_{\gamma},5^-_{\bar{\ell}},6^+_{\ell}) &=&  
\frac{\spbxb6.\gamma^{\mu}.(4+5).6}{t_{456} \spb4.5 \spb4.6} \;. 
\label{eq:demo3c}
\end{eqnarray}
Explicitly performing the calculation  we obtain the following tree level $\ell$-type amplitudes,
\begin{eqnarray}
\aa^{(0)}_{\ell}(1^+_q,2^-_{\bar{q}},3^+_g,4^+_{\gamma},5^-_{\bar{\ell}},6^+_{\ell}) 
& = & \tilde\aa_{\mu}(1^+_q,2^-_{\bar{q}},3^+_g) \, 
J^{\mu}_{\ell\bar{\ell}\gamma} (4^+_{\gamma},5^-_{\bar{\ell}},6^+_{\ell}) \nonumber \\
& = & i \, \frac{\spaxa2.(1+3).\gamma_{\mu}.2}{2 \spa1.3 \spa2.3} 
\, \frac{\spaxa5.\gamma^{\mu}.(4+6).5}{t_{456} \spa4.5 \spa4.6} \nonumber \\
& = & i \, \frac{\spa2.5^2}{\spa1.3 \spa2.3 \spa4.5 \spa4.6},\nonumber\\
\aa^{(0)}_{\ell}(1^+_q,2^-_{\bar{q}},3^+_g,4^-_{\gamma},5^-_{\bar{\ell}},6^+_{\ell}) 
& = & \tilde\aa_{\mu}(1^+_q,2^-_{\bar{q}},3^+_g) \, 
J^{\mu}_{\ell\bar{\ell}\gamma} (4^-_{\gamma},5^-_{\bar{\ell}},6^+_{\ell}) \nonumber \\
& = & i \, \frac{\spab2.(1+3).6 ^2 }{ t_{456}  \spa1.3 \spa2.3 \spb4.5 \spb4.6}. \nonumber
\end{eqnarray}
where we have used momentum conservation in order to simplify the results.

This procedure naturally extends both to other helicity amplitudes and to the NLO calculation. 
We have used the techniques described above to calculate the one-loop virtual amplitudes 
and real corrections to the $Z\gamma$+jet process. In addition, the $\ell$-type one-loop virtual amplitudes
have been cross-checked with an independent calculation using analytic unitarity techniques 
\cite{Britto:2004nc,Forde:2007mi,Mastrolia:2009dr}, utilizing the Mathematica package {\tt{S@M}} \cite{Maitre:2007jq}.
We present a more detailed breakdown of the calculation, as well as the full amplitudes, 
in Appendix~\ref{app:tree}.

\section{Calculation of $Z \gamma \gamma$ amplitudes}
\label{sec:zaacalc}

\begin{figure}
\begin{center}
\includegraphics[width=14cm]{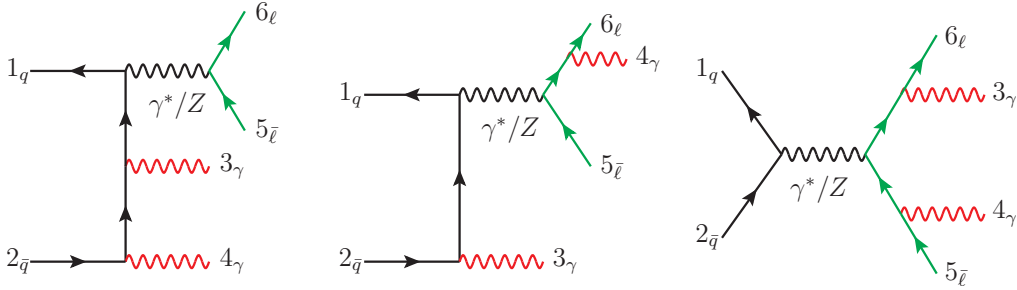}
\caption{Examples of leading order diagrams in the $Z \gamma \gamma$ process.
The three diagrams correspond to emission of both photons from the quark line (``$qq$-type'', left),
one photon from each of the quark and lepton lines (``$q\ell$-type'', centre) and
both photons from the lepton line (``$\ell\ell$-type'', right).
The particle labels correspond to momentum assignments that are all outgoing. }
\label{fig:zaalo}
\end{center}
\end{figure}
We now consider the NLO corrections to the triboson process $Z\gamma\gamma$, for which example leading order 
diagrams are shown in \fig~\ref{fig:zaalo}. NLO results exist in the literature both for this process~\cite{Bozzi:2011en} and $W\gamma\gamma$~\cite{Bozzi:2011wwa}, 
although this is the first time analytic results for this process have been written down. 
 Comparing to the $Z\gamma$+jet calculation presented in the previous section we observe 
that there is an increased number of distinct topologies related to the positioning of the two photons. The color structure
of this process is trivial compared to the $Z\gamma$+jet case, with only one such structure even at NLO.
Using the same notation as the previous section we define the following sub amplitudes corresponding to the 
cases where; 
\begin{itemize}
\item both photons are emitted from the quark line ($qq$-type),
\item one photon is emitted from each of the quark and lepton lines ($q\ell$-type),
\item both photons are emitted from the lepton line ($\ell\ell$-type).
\end{itemize}
Although there are nominally two $q\ell$ topologies it is clear they are related to each other by the exchange of the two photon momenta. 
Explicitly, the decomposition of the tree level $0 \rightarrow q \qb \gamma \gamma \bar\ell \ell$ 
amplitude into our sub-amplitudes listed above is,
\begin{eqnarray}
&& A^{(0)}(1_q,2_{\qb},3_{\gamma},4_{\gamma},5_{\bar{\ell}},6_{\ell}) = 4 e^4 \delta_{i_1 i_2} \nonumber \\
&& \times \bigg[ Q_q^2 \left(-Q_q + v_{L,R}^{q}v_{L,R}^{\ell} \prop{Z}(s_{56}) \right)
\aa^{(0)}_{qq}(1_q, 2_{\qb},3_{\gamma},4_{\gamma},5_{\bar{\ell}},6_{\ell}) \nonumber \\
&& \hspace{0.6cm} +Q_{\ell}Q_q \left(-Q_q + v_{L,R}^{q}v_{L,R}^{\ell} \prop{Z}(t_{456}) \right)
\aa^{(0)}_{q\ell}(1_q,2_{\qb},3_{\gamma},4_{\gamma},5_{\bar{\ell}},6_{\ell}) \nonumber \\
&& \hspace{0.6cm} +Q_{\ell}Q_q \left(-Q_q + v_{L,R}^{q}v_{L,R}^{\ell} \prop{Z}(t_{356}) \right)
\aa^{(0)}_{q\ell}(1_q,2_{\qb},4_{\gamma},3_{\gamma},5_{\bar{\ell}},6_{\ell}) \nonumber \\
&& \hspace{0.6cm}  +Q_{\ell}^2 \left(-Q_q + v_{L,R}^{q}v_{L,R}^{\ell} \prop{Z}(t_{3456}) \right)
\aa^{(0)}_{\ell\ell}(1_q,2_{\qb},3_{\gamma},4_{\gamma},5_{\bar{\ell}},6_{\ell})
\bigg]. 
\label{eq:zaacolor1}
\end{eqnarray}

The amplitudes for the $qq$-type diagrams are obtained from the subleading color contribution of the $q$-type 
diagrams in the $Z\gamma$+jet process, and similarly for the $q\ell$-type diagrams from the $\ell$-type $Z\gamma$+jet.
The amplitude for the $\ell\ell$-type diagrams may be obtained using the method that 
is described in Sec.~\ref{sec:zajcalc}, where we contract the $q\qb \rightarrow Z/ \gamma^*$ QCD current, 
$\tilde\aa_{\mu}(1_q,2_\qb)$ with the $Z/ \gamma^* \rightarrow \ell^+ \ell^- \gamma \gamma$ electroweak current.
However, in this paper we obtain the $\ell\ell$-type amplitudes simply from the $qq$-type amplitudes (for tree level)
 and from the $q\ell$-type (for tree level real emission) by crossing. The $\ell\ell$-type virtual amplitude
 is simply a vertex correction and is thus proportional to the $\ell\ell$-type tree level amplitude.
The explicit relations for the amplitudes are presented in Appendix~\ref{app:zgamgam}.


\section{Photon isolation and fragmentation}
\label{sec:photiso}

Final states containing photons can provide useful tests of the Standard Model. Experimental analyses
attempt to probe processes in which the photons directly participate in the hard scattering.
However, such studies are complicated by the additional production of photons through two
mechanisms. Firstly, photons can be  produced from the decays of unstable particles, for example
$\pi^0\rightarrow \gamma\gamma$.  Since these photons are not produced directly in the hard scattering
they are referred to as  secondary photons. The second category occurs due to the
fragmentation of QCD partons. Since the underlying production mechanism is a purely QCD
process, these fragmentation photons are copiously produced at hadron colliders. 

Secondary and fragmentation photons are usually associated with significant amounts of hadronic activity.
In order to reduce the effect of such photons, analyses typically
require that the amount of hadronic energy in the vicinity of the photon is limited, 
\begin{eqnarray}
\sum_{{\rm had} \in R_0} E_T^{\rm had} <   E_T^{\rm max}  \quad {\rm{with}} \quad R_0 = \sqrt{\Delta\phi^2+\Delta\eta^2} \;.
\label{expiso}
\end{eqnarray}
At the LHC, typical values for the parameters in Eq.~(\ref{expiso}) are a cone of size $R_0 \sim 0.4$
and a limit on the maximum transverse energy, $E_T^{\rm max} \sim 5$ GeV.  
This requirement is referred to as an isolation cut.

Imposing such an isolation procedure raises an additional complication on the theoretical side.
Consider the production of a photon in association with a jet. At LO the matrix element is finite since
the  jet and photon cuts require that the two particles be well-separated in phase space. However, at NLO
bremsstrahlung diagrams occur that involve a photon and two QCD partons. In addition to the usual QCD infrared 
singularities, which cancel in the combination with the virtual corrections, these diagrams contain a collinear singularity 
corresponding to photon emission from a quark. Since we require the photon to be resolved, this singularity has no virtual counterpart.
Further, attempting to remove this singularity by requiring that no QCD radiation is present in the cone around the photon removes 
a region of phase space for soft gluon radiation, thus rendering the calculation infrared unsafe. 

One approach to avoid this problem is to isolate the photon in a different manner to
that described above. An alternative isolation procedure, proposed by Frixione~\cite{Frixione:1998jh},
allows arbitrarily soft radiation in the cone, whilst still eliminating the collinear fragmentation
pole. The isolation criterion is, 
\begin{eqnarray} 
\sum_{{\rm had}}  E_T^{\rm had} \, \theta(R-R_{{\rm had},\gamma})
 < \epsilon_h E_T^{\gamma}\bigg(\frac{1-\cos{R}}{1-\cos{R_0}}\bigg)^{n} 
 \qquad \mbox{for all}~R \le R_0 \;,
\label{eq:Frixiso}
\end{eqnarray}
where $n$ and $\epsilon_h$ are parameters of the algorithm. 
This prescription has the theoretical advantage that the fragmentation contributions do not have to be
considered. On the other hand, this ``smooth cone'' isolation is difficult to apply
experimentally.

Alternatively, one can return to the definition of isolation that is used in the experiments,
Eq.~(\ref{expiso}), and regularize the collinear fragmentation pole. This is achieved by absorbing the
collinear splitting into the kernel of a fragmentation function that behaves in a manner analogous to
initial state PDFs~\cite{Bourhis:1997yu}. These fragmentation functions, which satisfy the DGLAP equation,
must be extracted from data due to their non-perturbative nature. The NLO prediction then consists of the
usual NLO diagrams plus a QCD LO matrix element coupled to the relevant fragmentation function,
\begin{eqnarray} 
\sigma^{\gamma}_{NLO} = \sigma_{NLO,{\rm direct}}^{\gamma}(M_F)
+\int_0^1 d z \sum_a \sigma^{a}_{f} D_{a\rightarrow\gamma}(z,M_F) \;.
\label{eq:fragform}
\end{eqnarray} 
In this equation $\sigma^a_f$ represents the production of a final state parton of species $a$ in place of the final
state photon, and $D_{a\rightarrow \gamma}(z,M_F)$ is the fragmentation function with $p_\gamma=z p_a$. 
Note that the introduction of the fragmentation functions also requires the addition of a new scale,
the fragmentation scale $M_F$. As a result the separation between the direct and fragmentation
pieces is not unique and only the sum is theoretically well defined. In MCFM we have implemented the
fragmentation sets of BFG~\cite{Bourhis:1997yu} and GdRG~\cite{GehrmannDeRidder:1998ba}. 

\section{Phenomenology}
\label{sec:zaj_ph}


We have included the processes described in sections~\ref{sec:zajcalc} and ~\ref{sec:zaacalc} into the NLO parton level code 
MCFM which is available publicly~\cite{MCFMweb}. This builds on the existing $W\gamma$ and $Z\gamma$ 
processes already in the code. MCFM uses the dipole subtraction scheme formulated by Catani and
Seymour~\cite{Catani:1996vz} in order to isolate singularities in the  
virtual and real contributions. We use the following MCFM default 
electroweak (EW) parameters in our calculation, 
\begin{eqnarray}
M_Z = 91.1876 \,\,{\rm{GeV}} \;, \quad M_W = 80.398  \,\,{\rm{GeV}} \;,\nonumber\\
\Gamma_Z = 2.4952  \,\,{\rm{GeV}} \;, \quad \Gamma_W = 2.1054 \,\,{\rm{GeV}} \;, \nonumber\\
G_F=0.116639  \times 10^{-4}  \,\,{\rm{GeV^{-2}}} \;, \quad m_t = 172.5 \,\,{\rm{GeV}} \;. \nonumber
\end{eqnarray}
The remaining EW parameters are defined using the above as input parameters.
In our calculations we use
the CTEQ6L1 PDF set at LO, the CT10 PDF set at NLO~\cite{Lai:2010vv}
and Set II of BFG~\cite{Bourhis:1997yu} for the fragmentation functions. 

\subsection{$Z\gamma$ + jet at the LHC} 

In this section we investigate the phenomenology of the production of $Z\gamma$ and an associated jet at LHC operating energies.  For typical experimental photon
selection cuts (with $p_T^\gamma >15$~GeV), the  $e^+e^-\gamma$ inclusive cross section is relatively large, about $1$~pb at 7 TeV~\cite{Campbell:2011bn}. As a
result, the process in which an additional hard jet is radiated can also be readily observed.

\begin{figure}[t]
\begin{center}
\includegraphics[scale=1.2]{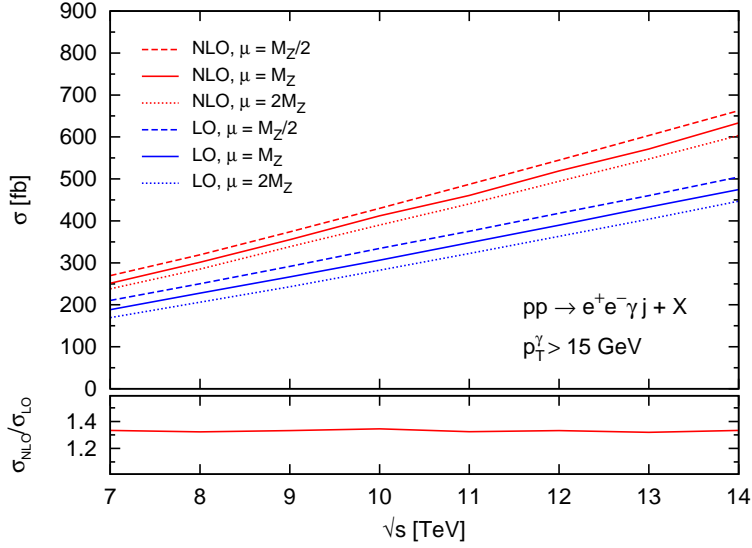}
\end{center}
\caption{Dependence of the LO (blue) and NLO (red) inclusive cross section of 
$pp \rightarrow \ell^+\ell^-\gamma$+jet process on the renormalization/factorization scale, for a range of LHC operating energies}
\label{fig:xs_sqrts}
\end{figure}

We therefore begin by presenting $\ell\bar\ell\gamma$+jet cross sections for the LHC at a range of operating energies. We base our selection cuts 
on those used in the most recent ATLAS analysis~\cite{Aad:2012mr}. 
In order to pass the selection cuts an event must satisfy the following criteria, 
\begin{eqnarray}
\mathrm{Photon}   & : &  |\eta_{\gamma}| < 2.37, \; R_{\ell \gamma} > 0.7 \;,\nonumber \\
&& \begin{array}{ll}
p_T^{\gamma} > 15~\mathrm{GeV} &  {\rm{(low-}} p_T), \\
p_T^{\gamma} > 60~\mathrm{GeV} &  {\rm{(intermediate-}} p_T), \\
p_T^{\gamma} > 100~\mathrm{GeV} &  {\rm{(high-}} p_T). \\
\end{array}
\nonumber\\
\mathrm{Leptons}   & : &  m_{\ell^+ \ell^-} > 40 \; \mathrm{GeV}, \; p_T^{\ell} > 25 \;
 \mathrm{GeV}, \; |\eta_{\ell}| < 2.47, \; R_{\ell j} > 0.6, \nonumber \\
& & E_T^{\mathrm{miss}} > 25 \; \mathrm{GeV} \;. \nonumber\\
\mathrm{Jets}     & : &  p_T^{j} > 30\;  \mathrm{GeV}, \; |y_j| < 4.4, \; \;  k_T \; \mathrm{algorithm \; with} \; R=0.4 \;.
\label{eq:cutsdefn}
\end{eqnarray}
In addition photons are isolated by requiring that $E_T^{\rm had} \in R_0 < E_T^{\rm max}$ with $R_0=0.3$ and $E_T^{\rm max} = 6$ GeV.
LO and NLO cross sections under the fiducial cuts listed above, for operating energies of 7 and 8 TeV, are shown in Table~\ref{table:xsection}.
The factorization, renormalization and fragmentation scales are set equal to one other, $\mu_F = \mu_R = M_F = M_Z $. 
Note that we have not provided predictions for the neutrino case with a low photon cut since it may not be possible to trigger on such events.
First, we see that the effect of tightening the photon cuts from $15$ to $60$~GeV is to lose about an order of magnitude in the yield. Raising the cut even further, to $100$~GeV
reduces the cross sections by a further factor of three.
In all cases the effect of the NLO corrections is to increase the cross section by a factor of about 1.3, independent of the operating energy.
For the $e^+e^-\gamma$+jet final state using the low photon $p_T$ cut, we extend the dependence on operating energy up to the LHC design target of 14 TeV in Fig.~\ref{fig:xs_sqrts}. 
As is clear from the figure, the $K$-factor is constant across the entire foreseeable range of LHC operating energies.
At 14~TeV the cross section is twice as large as at 7~TeV but the fractional uncertainty is approximately the same.
\renewcommand{\baselinestretch}{1.6}
\begin{table*}[t]
\begin{center}
\begin{tabular}{|c|cr|c|c|}
\hline
\hline
Photon Cut & Cross Section &  & $\sqrt{s}=$ 7 TeV &  $\sqrt{s}=$ 8 TeV \\ \hline
 \multirow{2}{3cm}{\centering low-$p_T$ \\ ($p_T^{\gamma}>15$~GeV)} 
                            & \multirow{2}{3cm}{$e^+e^- \gamma +\mbox{jet} + X$} & LO & $ 188^{+ 11.6 \%}_{-10.0 \%}$ & $ 228^{+ 9.9 \%}_{- 9.4 \%}$ \\
                            &  & NLO & $ 252^{+ 7.4 \%}_{- 5.2 \%}$ & $ 301^{+ 5.9 \%}_{- 5.4 \%}$ \\ \hline \hline
 \multirow{4}{3cm}{\centering intermediate-$p_T$ \\ ($p_T^{\gamma}>60$~GeV)  } 
                            & \multirow{2}{3cm}{$e^+e^- \gamma +\mbox{jet} + X$} & LO & $ 18.6^{+ 16.5 \%}_{- 13.3 \%}$ & $ 23.0^{+ 15.4 \%}_{-12.6 \%}$ \\
                            &  & NLO & $ 24.6^{+ 8.1 \%}_{- 6.8 \%}$ & $ 30.7^{+ 6.9 \%}_{-6.8 \%}$ \\
			    \cline{2-5}
                            & \multirow{2}{3cm}{$3(\nu \bar{\nu}) \gamma +\mbox{jet} + X$} & LO & $ 182^{+ 16.8 \%}_{-13.5 \%}$ & $ 229^{+ 15.7 \%}_{- 12.8 \%}$ \\
                            &  & NLO & $ 249^{+ 8.9 \%}_{-8.1 \%}$ & $ 316^{+ 9.2 \%}_{- 7.9 \%}$ \\ \hline \hline
 \multirow{4}{3cm}{\centering high-$p_T$ \\ ($p_T^{\gamma}>100$~GeV)  } 
                            & \multirow{2}{3cm}{$e^+e^- \gamma +\mbox{jet} + X$} & LO & $ 6.70^{+ 19.0 \%}_{- 14.9\%}$ & $ 8.52^{+ 17.9 \%}_{- 14.4 \%}$ \\
                            & & NLO & $ 8.34^{+ 6.6 \%}_{- 6.1 \%}$ & $ 10.7^{+ 6.0 \%}_{-6.1 \%}$ \\
			    \cline{2-5}
                            & \multirow{2}{3cm}{$3(\nu \bar{\nu}) \gamma +\mbox{jet} + X$} & LO & $ 65.4^{+19.1  \%}_{-15.0  \%}$ & $ 84.6^{+ 18.0 \%}_{- 14.2 \%}$ \\
                            &  & NLO & $ 86.1^{+8.5  \%}_{- 8.3 \%}$ & $ 112^{+ 8.9 \%}_{-7.8 \%}$ \\
\hline\hline
\end{tabular}
\renewcommand{\baselinestretch}{1.0}
\caption{LO and NLO cross sections (in femtobarns) for $e^+ e^- \gamma$+jet and $\nu \bar\nu \gamma$+jet production at the LHC
with centre-of-mass energy 7 and 8 TeV. Note that, for the neutrino case, we have summed over three flavours of neutrino.
The quoted value represents the scale choice of $M_Z$ with the 
percentages indicating the shift in the central value upon varying this scale by a factor of two. }
\label{table:xsection}
\end{center}
\end{table*}
\renewcommand{\baselinestretch}{1.0}

In this section we have focused on NLO results for inclusive quantities. In the following section we will more closely follow the experimental
setup by splitting the analysis into jet bins.

\subsection{$Z\gamma$ and $Z\gamma+$ jet :  Exclusive predictions} 

In this section we will investigate more exclusive quantities involving a final state consisting of $Z\gamma$ and a fixed number of jets. As we discussed in the introduction,
previous versions of MCFM~\cite{Campbell:2011bn} are able to predict quantities accurate to NLO in the 0-jet bin, and to LO in the 1-jet bin. Using the
calculations presented in this paper we are able to extend the 1-jet bin to NLO accuracy. With the ATLAS results in hand~\cite{Aad:2012mr} we will 
also take the opportunity to re-evaluate predictions for the 0-jet bin and reassess their theoretical uncertainties.

The ATLAS paper~\cite{Aad:2012mr} studied the production of $Z\gamma$ final states, separating events based upon the number 
of reconstructed jets. Using this approach they were able to provide measurements of both the inclusive  i.e. $\sigma(V\gamma + X)$ and exclusive i.e. $\sigma(V\gamma + 0~\mbox{jet} )$
cross sections. 
Explicitly, the results for the $\ell^+\ell^-\gamma$ final state (averaged over electron and muon channels) quoted in Ref.~\cite{Aad:2012mr} are:
\begin{center}
Low$-p_T$ region : 
\begin{eqnarray}
{\rm Inclusive:} \quad
\sigma^{ATLAS}=1.29 \pm 0.05 \pm 0.15 \,\rm{pb} \;, && \quad \sigma^{MCFM}=1.22\pm0.05 \,\rm{pb} \nonumber \\
{\rm Exclusive:} \quad
\sigma^{ATLAS}=1.05 \pm 0.04 \pm 0.12 \,\rm{pb} \;, && \quad \sigma^{MCFM}=1.03 \pm 0.04 \,\rm{pb} \nonumber
\label{eq:data_theory_xs_low}
\end{eqnarray}
Intermediate$-p_T$ region: 
\begin{eqnarray}
{\rm Inclusive:} \quad
\sigma^{ATLAS} =68 \pm 8\pm  5 \,\rm{fb} \;, && \quad \sigma^{MCFM}=58\pm 5 \,\rm{fb} \nonumber  \\
{\rm Exclusive:} \quad
\sigma^{ATLAS}= 47 \pm 7 \pm 4 \,\rm{fb}  \;, && \quad \sigma^{MCFM}=40 \pm 3 \,\rm{fb}
\label{eq:data_theory_xs_int}
\end{eqnarray} 
\end{center}
The experimental measurement is given with the statistical (first) and systematic (second) errors separately. The MCFM prediction is taken from
the parton level values quoted in the paper (the collaboration also presents a particle-level corrected result). 
We see that, although the low-$p_T$ results are in agreement with the MCFM prediction, the intermediate-$p_T$ measurements are somewhat
higher, albeit with larger uncertainties. A further feature is also apparent: the relative theoretical uncertainty on 
the exclusive cross sections is comparable, or smaller, than that on the inclusive prediction.
Since the exclusive calculation introduces a new scale, corresponding to the transverse momentum of any additional jets that are
effectively vetoed, one expects the calculation to have a richer structure and hence a larger theoretical uncertainty.
The principle aim of this section is thus to investigate an improved method for estimating the scale uncertainty~\cite{Stewart:2011cf} in order
to provide updated theoretical results for the 0-jet bin. In addition we will provide new theoretical predictions for the 1-jet bin that can be compared against future measurements.

\begin{figure}[t]
\begin{center}
\includegraphics[scale=1.15]{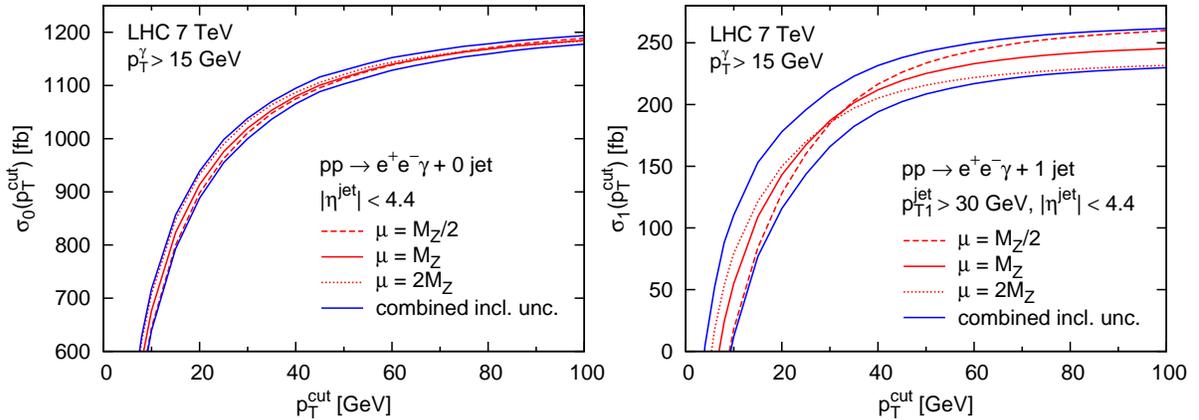}
\caption{A comparison of estimating scale uncertainties using the standard method of varying around a central scale by a factor of two (red) with 
the ST method (blue). The LHC operating energy is $7$~TeV and the photon cut takes the low value, i.e. $p_T^\gamma>15$~GeV. The left 
hand plot shows the error estimation at NLO in the exclusive 0-jet bin. The plot on the right side represents the same quantity for the 
NLO 1-jet exclusive cross section.}
\label{fig:ST_15_7}
\end{center}
\end{figure}
The naive method of estimating theoretical uncertainties by scale variation, which is of course only a crude
estimate of missing higher order contributions, has been shown to be extremely dangerous in the presence of jet vetoes~\cite{Stewart:2011cf}. 
The authors or Ref.~\cite{Stewart:2011cf} propose that instead of using the scale variation of the exclusive cross section as a measure of the
uncertainty one should use the following,
\begin{eqnarray}
\Delta^2_{Z\gamma}= \Delta^2_{\ge Z\gamma} + \Delta^2_{\ge Z\gamma j} \;.
\end{eqnarray}
In this equation $\Delta_{Z\gamma}$ represents the total uncertainty in the exclusive 0-jet bin and
 $\Delta_{\ge Z\gamma}$ and  $\Delta_{\ge Z\gamma j}$ represent the uncertainties obtained from the inclusive calculation 
of $Z\gamma$ and $Z\gamma$+jet respectively.
In this way the two perturbation series in $\alpha_S$ which make up the exclusive prediction are treated as uncorrelated. This ensures that no accidental cancellation between the scale-dependent 
coefficients in the perturbation series occurs. As is stressed in Ref.~\cite{Stewart:2011cf}, if $\Delta_{ \ge Z\gamma}$
is calculated  at NLO then $\Delta_{\ge Z\gamma j}$ should be calculated at LO.
This makes sense since, although we have computed $\Delta_{\ge Z\gamma j}$  at NLO, we should not expect improvement in the errors in the 0-jet bin without first calculating the NNLO
corrections.

\begin{figure}[t]
\begin{center}
\includegraphics[scale=1.15]{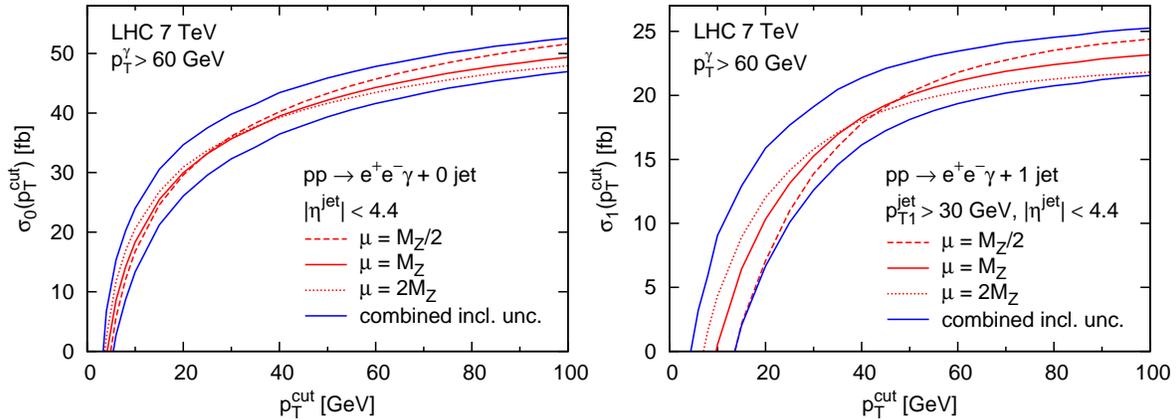}
\caption{The same as Figure~\protect\ref{fig:ST_15_7} but for the intermediate photon cut,  i.e. $p_T^\gamma>60$~GeV.}
\label{fig:ST_60_7}
\end{center}
\end{figure}
In Figs.~\ref{fig:ST_15_7} and~\ref{fig:ST_60_7} we present the dependence of the exclusive cross sections on the jet veto,
for the 0- and 1-jet bins at NLO. We vary the scales using two
different techniques. Firstly we use the more traditional approach of choosing a central scale (in this case $M_Z$) and  varying it by a factor of two in each direction.
Secondly we vary the scales using the approach discussed above (hereafter referred to as the ST method),
in which the 0- and 1-jet errors are defined as follows, 
\begin{eqnarray}
\left(\Delta_{Z\gamma}^{NLO}\right)^2 &=& \left(\Delta_{\ge Z\gamma}^{NLO}\right)^2 + \left(\Delta_{\ge Z\gamma j}^{LO}\right)^2 \;, \\
\left(\Delta_{Z\gamma j}^{NLO}\right)^2 &=& \left(\Delta_{\ge Z\gamma j}^{NLO}\right)^2 + \left(\Delta_{\ge Z\gamma jj}^{LO}\right)^2 \;. 
\end{eqnarray}
We calculate the uncertainties thus obtained for a range of veto scales. Note that,  for the 1-jet bin, we always require  at least one jet with $p_T > 30$ GeV and then vary the veto parameter for
the second jet.  It is clear from the figures that the ST method provides a more realistic measure of theoretical uncertainty than simply using the usual method of scale variation. The ST method has
the pleasing feature  of reproducing the inclusive error in the large-veto limit and, in addition, never results in a value of the jet veto for which the uncertainty vanishes.
For example, this is the case for the $Z\gamma$+jet calculation with $p_T^\gamma >15$~GeV (Fig.~\ref{fig:ST_15_7}, right) when using a jet veto of around $30$~GeV and the
usual method of scale variation.

Both figures also illustrate that the $Z\gamma$+jet predictions have a larger ST scale variation than $Z\gamma$. This can be explained by considering the differences between the Born production mechanisms in both. 
For the 0-jet bin the Born production process is purely electroweak and as a result the scale dependence is minimal, resulting from the factorization scale used in the PDFs. Clearly the dependence of 
the 1-jet bin on $\alpha_S$ occurs naturally at LO, thus yielding a much stronger dependence on $\alpha_S(\mu_R)$ than the 0-jet bin.

\subsection{ $Z\gamma$ and $Z\gamma +$ jet : Cross section summary} 
To conclude this section we present the NLO predictions for the 0- and 1- jet exclusive and inclusive cross sections. We use the ST method to estimate scale uncertainty and
also include uncertainties due to the PDFs and fragmentation contributions. 
The PDF uncertainties are obtained by using the $68$\% confidence level sets of CT10~\cite{Lai:2010vv}. 
In order to estimate the uncertainty arising from the fragmentation contributions we re-calculate the cross section 
using the fragmentation functions of Ref.~\cite{GehrmannDeRidder:1998ba} and compare the results with our default 
fragmentation set. This provides a crude estimate of the uncertainty arising from the modelling of the non-perturbative pieces 
of the fragmentation functions. 
These results are collected in Tables~\ref{table:xs_ST_zg} and~\ref{table:xs_ST_zgj}.
Note that imposing a jet-lepton separation (c.f. Eq.~(\ref{eq:cutsdefn})) means that the inclusive $Z\gamma$ cross sections presented in Table~\ref{table:xs_ST_zg} depend on the jet definition.

From these tables we can read off our theoretical predictions for the $e^+e^-\gamma$ cross sections in the $0$- and $1$-jet bins,
using the fiducial cuts employed in the ATLAS study~\cite{Aad:2012mr}. Converting from relative to absolute uncertainties and adding them linearly we have:
\begin{eqnarray*}
0-{\rm jet:} &&
\sigma^{NLO}(p_T^\gamma > 15~{\rm GeV}) = 1.02 \pm 0.06  \,{\rm pb} \;,
\qquad \sigma^{NLO}(p_T^\gamma > 60~{\rm GeV}) = 35.7^{+6.4 }_{-5.6 } \,{\rm fb} \\
1-{\rm jet:} &&
\sigma^{NLO}(p_T^\gamma > 15~{\rm GeV}) = 187^{+30}_{-27} \,{\rm fb}\;, 
\qquad \qquad \sigma^{NLO}(p_T^\gamma > 60~{\rm GeV}) = 15.3^{+4.8}_{-3.6} \,{\rm fb} 
\label{eq:best0and1preds}
\end{eqnarray*} 
In the 0-jet bin we see, by comparing with the predictions quoted in Eq.~(\ref{eq:data_theory_xs_int}), that
our revised results are similar to those presented in Ref.~\cite{Aad:2012mr}, with some of the difference attributable to the
different choice of PDF set. However, 
the uncertainties are 50\% larger due to the different treatment of the scale
uncertainty. In the 1-jet bin, the combined theoretical uncertainty is very large, indicating that a comparison with the theoretical prediction is of
questionable value. In contrast the uncertainty on the inclusive $1$-jet prediction (upper rows of Table~\ref{table:xs_ST_zgj}) is much smaller, at the level of $10$\%,
so that a much more meaningful
comparison could be made. In passing, we note that our predictions for the 1-jet inclusive cross section are compatible with the difference between the
inclusive and exclusive $0$-jet bin $Z\gamma$ results presented in Ref.~\cite{Aad:2012mr}.
\renewcommand{\baselinestretch}{1.6}
\begin{table}[t]
\begin{center}
\begin{tabular}{|c|c|}
\hline
\hline
 & Inclusive NLO $(e^+e^-\gamma + X )$ [pb]  \\
 \hline
  {$p_T^{\gamma} >$ 15 GeV} &  ~~~$ 1.21 ^{+0.7  \%}_{-0.5 \%}$(scale)$\pm 3.5 \% $(PDF)$ \pm 0.4 \%$(frag)  \\
  {$p_T^{\gamma} >$ 60 GeV} &  $ 0.0545 ^{+ 5.7\%}_{-4.1  \%}$(scale)$ \pm 3.7  \% $(PDF)$ \pm 1.2  \%$(frag)  \\
  \hline
\hline
 & Exclusive NLO $(e^+e^-\gamma + \mbox{no jets} )$ [pb]  \\
 \hline
  {$p_T^{\gamma} >$ 15 GeV} &  ~~$ 1.02^{+1.8  \%}_{-1.9 \%}$(scale)$ \pm 3.8 \% $(PDF)$ \pm 0.5  \%$(frag)  \\
  {$p_T^{\gamma} >$ 60 GeV} &  $ 0.0357 ^{+11.6  \%}_{-9.4 \%}$(scale)$ \pm 4.4 \% $(PDF)$ \pm 1.8 \%$(frag) \\  
\hline\hline
\end{tabular}
\renewcommand{\baselinestretch}{1.0}
\caption{NLO (inclusive and exclusive) predictions for the $Z\gamma$ cross sections at LHC with
 $\sqrt{s}$ = 7 TeV. The NLO exclusive cross section, containing no identified jets, is defined
 using our usual jet cuts in Eq.~(\protect\ref{eq:cutsdefn}).}
\label{table:xs_ST_zg}
\end{center}
\end{table}
\renewcommand{\baselinestretch}{1.0}
\renewcommand{\baselinestretch}{1.6}
\begin{table}[t]
\begin{center}
\begin{tabular}{|c|c|}
\hline
\hline
 & Inclusive 1-jet NLO $(e^+e^-\gamma + \mbox{jet} +  X )$ [fb]  \\
 \hline
  {$p_T^{\gamma} >$ 15 GeV} &  $252^{+ 7.4 \%}_{- 5.2 \%}$(scale)$ \pm 2.0 \%$(PDF)$ \pm 0.6 \%$(frag)  \\
    {$p_T^{\gamma} >$ 60 GeV} &  $24.6^{+ 8.1 \%}_{- 6.8 \%}$(scale)$ \pm 2.5 \%$(PDF)$ \pm 1.9 \%$(frag)  \\
  \hline
\hline
 & Exclusive 1-jet NLO $(e^+e^-\gamma + \mbox{jet} )$ [fb]  \\
 \hline
  {$p_T^{\gamma} >$ 15 GeV} &  $188^{+ 13.0\%}_{- 11.2 \%}$(scale)$ \pm 2.4 \%$(PDF)$ \pm 0.8 \%$(frag)  \\
  {$p_T^{\gamma} >$ 60 GeV} &  $15.3^{+25.3 \%}_{- 17.4 \%}$(scale)$ \pm 2.9 \%$(PDF)$ \pm 3.1 \%$(frag) \\  
\hline\hline
\end{tabular}
\renewcommand{\baselinestretch}{1.0}
\caption{NLO (inclusive and exclusive) predictions for the $Z\gamma$ + jet cross sections at LHC with
 $\sqrt{s}$ = 7 TeV. Jets are defined
 using our usual jet cuts in Eq.~(\protect\ref{eq:cutsdefn}).}
\label{table:xs_ST_zgj}
\end{center}
\end{table}
\renewcommand{\baselinestretch}{1.0}

\begin{figure}[h]
\begin{center}
\includegraphics[scale=1.5]{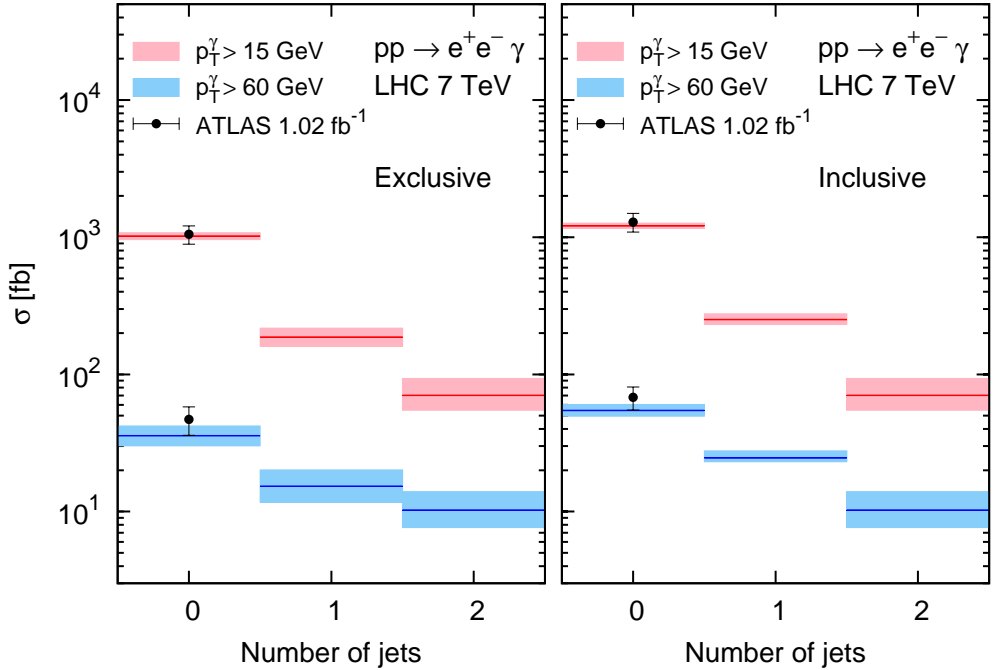}
\caption{A summary of $e^+e^-\gamma$ cross sections obtained using MCFM. The left panel shows the exclusive predictions, i.e. for
a specific number of jets and the right panel depicts the inclusive predictions, i.e. for at least the given number of jets. The ATLAS
data, for the $0$-jet bin in each case, are taken from Ref.~\protect\cite{Aad:2012mr}. \label{fig:xsecsummary}}
\end{center}
\end{figure}

A summary of our parton-level predictions, together with the ATLAS results from Ref.~\cite{Aad:2012mr}, is shown in Fig.~\ref{fig:xsecsummary}.
Note that all predictions are accurate to NLO, except for the $2$-jet results that are purely LO and therefore identical in the
exclusive and inclusive cases. The uncertainty on the LO results corresponds only to scale variation, which is already considerable.

\subsection{Photon $p_T$ spectrum in $Z\gamma$ and $Z\gamma$+jet}
Next we consider the $p_T$ spectrum of the photon produced in events containing $\ell\bar{\ell}\gamma$ and either zero or one jet. This is an important kinematic distribution
since any deviation from the expected SM prediction may indicate the presence of anomalous couplings between  the gauge bosons. In addition, searches in the
missing $E_T$+photon channel use
the photon as a probe in order to search for the production of dark matter at colliders \cite{Chatrchyan:2012tea}.   In both cases accurate modelling of the background is essential in order
to constrain, or observe, the new physics.

\begin{figure}[t]
\begin{center}
\includegraphics[scale=1.3]{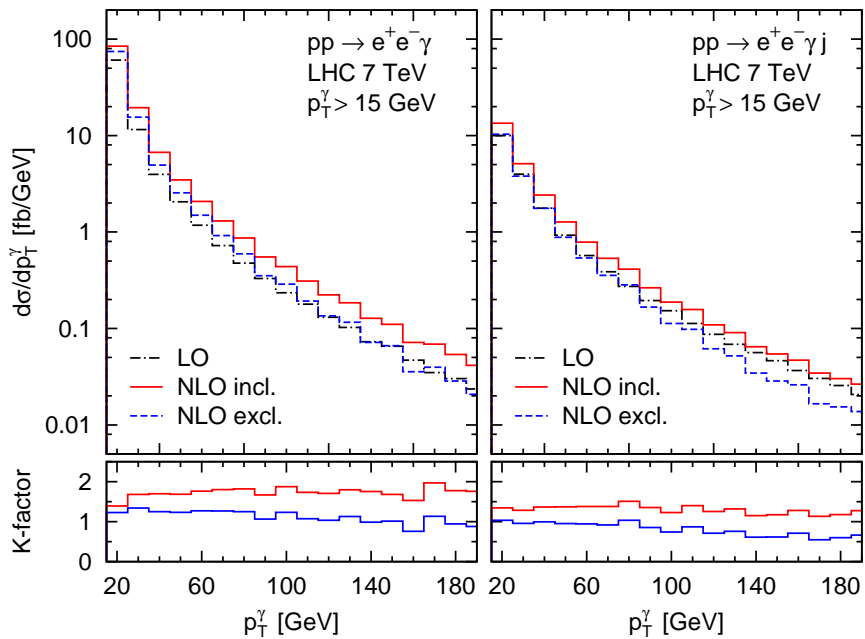}
\caption{The upper panel presents the photon $p_T$ spectrum using the low $p_T$ photon selection requirements at 7 TeV. Shown 
are the LO predictions (black dot-dashed), NLO exclusive prediction (blue-dashed) and NLO inclusive predictions (red-solid). The left hand plot 
shows the photon $p_T$ for $Z\gamma$, whilst the right hand plot is the spectrum from $Z\gamma$+jet. The lower panel illustrates the inclusive (red) and 
exclusive (blue) $K$-factor for both predictions. \label{fig:pt_7_15}}
\end{center}
\end{figure}

Therefore in Fig.~\ref{fig:pt_7_15} we present the photon $p_T$ spectrum for our low photon $p_T$ selection requirement. We observe that in the
region $15$--$200$ GeV the inclusive $K$-factor  is relatively stable in both the 0- and 1-jet bins. In the 0-jet case the $K$ factor increases gently as the $p_T$ grows. However, for
higher $p_T$ values it approaches 2. The increasing $K$-factor is hardly surprising since, as discussed in~\cite{Campbell:2011bn}, at NLO one has large corrections to the inclusive rate from diagrams
that include a gluon in the initial state, a high-$p_T$ jet and a relatively soft $Z$. Once this new kinematic regime is stifled by the application of a veto, the $K$ factor reduces and
becomes flatter, since the allowed kinematic region is more similar to the LO one.  The inclusive $K$-factor for the 1-jet bin is smaller than its corresponding 0-jet counterpart. This is
primarily because the presence of a jet in the Born topology  allows more of the phase space for the $Z$ and the photon to be explored at LO. In addition the presence of a gluon in the
initial state at LO results in a more modest increase than the 0-jet bin since there is  no large PDF enhancement of the real corrections. For the 1-jet bin the exclusive $K$-factor drops
below 1 and falls significantly in the higher $p_T$ region. We note that, for transverse momenta beyond the ranges that are presented here, our central scale choice of $M_Z$ may
no longer be appropriate and an event-by-event scale such as $\sqrt{M_Z^2 + (p_T^\gamma)^2}$ may be more reliable.

It is interesting to consider the ratio of photon $p_T$ spectra that could be constructed from a measurement of final states containing  $\ell^+\ell^-\gamma$ and
$\nu\bar{\nu}\gamma$. This ratio may be useful for several reasons. Firstly the basic interpretation in the SM is that this ratio is sensitive to the
contributions in which a photon is radiated from  the final state leptons. Secondly, the ratio should suffer from fewer experimental ambiguities since one expects some
cancellation of systematic errors. Thirdly, the ratio could be sensitive to models that modify the spectrum in only one of the channels.
For instance, in dark matter scenarios only the photon spectrum in the missing $E_T$ + photon channel is modified. In contrast, in the case of anomalous couplings one would expect both $p_T$ spectra
to be altered in the same way so that the ratio is the same as in the SM.

\begin{figure}[t]
\begin{center}
\includegraphics[scale=1.3]{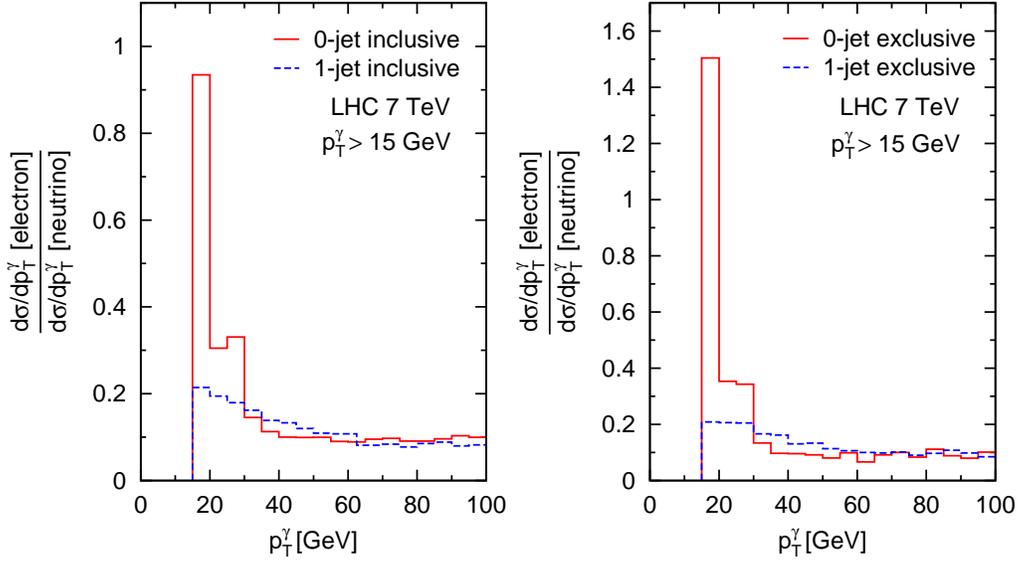}
\caption{The ratio of the photon transverse momentum spectra in charged lepton and neutrino final states
for inclusive $Z\gamma$ and $Z\gamma$+jet production (left) and exclusive $Z\gamma$ + 0-jet and $Z\gamma$ + 1-jet production (right).
\label{fig:ptgamratio}}
\end{center}
\end{figure}

In Fig.~\ref{fig:ptgamratio} we present NLO predictions for such ratios in the presence of either $0$ or $1$ jets. In constructing these
ratios we have considered a single lepton flavor but all three species of neutrinos.
We observe that the $0-$jet ratio has a strong peak in the low-$p_T$ region associated with the radiation  of softer photons from final state leptons. This
is somewhat diminished in the $1-$jet case, where there is an additional source of soft photons from events in which the $Z$ boson almost balances with a hard jet.
The tails in both cases tend to a constant ratio. We note that this constant is lower than might be expected from the relative branching ratios ($\sim 1/6$), due to 
the difference in selection criteria for each process. For the invisible $Z$ decay only the total neutrino momentum is subject to a $p_T$ cut and the rapidity
is unconstrained, cuts that are less restrictive than those in the electron channel. 

\subsection{$Z\gamma\gamma$ phenomenology}
\label{sec:zaa_ph}

 In this section we present some phenomenological studies for the final states $\ell^+\ell^-\gamma\gamma$ and
 $\nu\bar\nu\gamma\gamma$ at the LHC.  Although the production cross section for $Z\gamma$ is large the cost of radiating a further
 electroweak boson is severe, resulting in small rates at  hadron colliders. As a result the observation of the $Z\gamma\gamma$
 cross section remains an experimental challenge. The process $Z\gamma\gamma$ has received theoretical  attention in the past, with
 NLO corrections calculated in Ref.~\cite{Bozzi:2011en}. However our study is the first to include the effects of photon
 fragmentation, allowing for the isolation of photons in a manner that is similar to that performed in experiments. We have checked
 our results against those presented in~\cite{Bozzi:2011en}, using the Frixione isolation procedure described by Eq.~(\ref{eq:Frixiso}),
 and find agreement within Monte Carlo uncertainties. 

 Our aim in this section is to describe the phenomenology of the $Z\gamma\gamma$ process, focussing primarily on the LHC operating at 8
 TeV. Since the 8 TeV data set from 2012 alone could be around  20 fb$^{-1}$ per experiment and the cross section for the
 $e^+e^-\gamma\gamma$ process is about $2$~fb, analyses using this data set should have
 the best chance of observing this small SM process. The successful observation of this process at a hadron collider  will at the
 very least instill further confidence in the ability of the LHC to identify rare processes with such small cross sections. In addition these rare SM
 processes may yield new insights into physics beyond the Standard Model.  Observing a significantly different total rate than that
 which is predicted by the SM could be a sign of new physics. For this reason it is crucial to have predictions for the total rate
 accurate to NLO. 
 
\begin{figure}
\begin{center} 
\includegraphics[scale=1.4,angle=0]{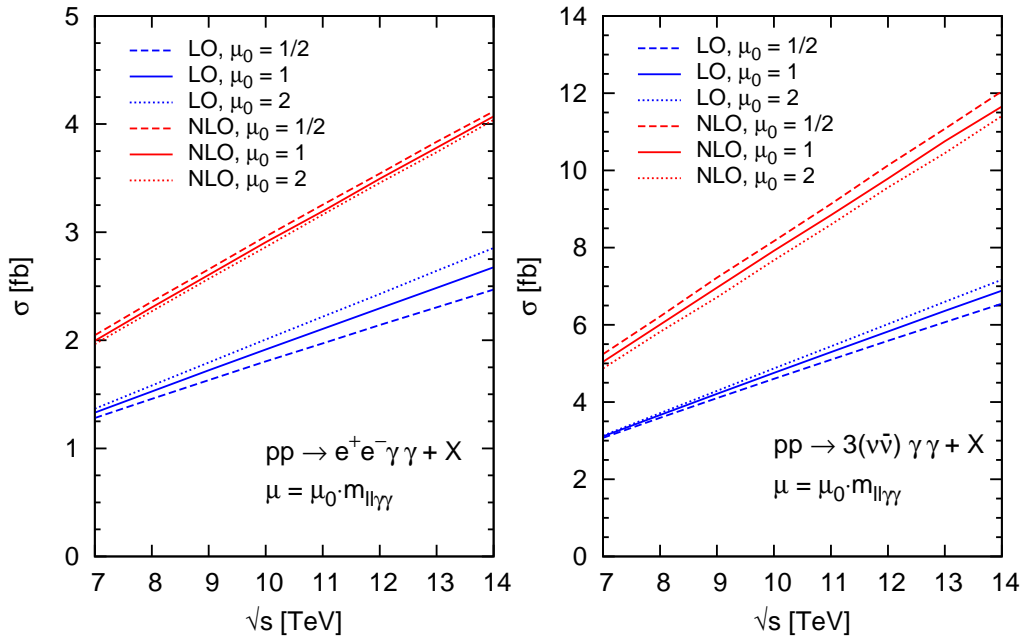} 
\caption{Dependence of the LO (blue) and NLO (red) inclusive cross section 
of $pp \rightarrow \ell^+ \ell^- \gamma \gamma$ (left) and 
$pp \rightarrow 3(\nu \bar{\nu}) \gamma \gamma$ (right) processes
on the renormalization/factorization scale, 
for a range of LHC operating energies.}
\label{fig:gamgam_xs}
\end{center}
\end{figure}

 We present cross sections for the production of $\ell^+\ell^-\gamma\gamma$ and $\nu\overline{\nu}\gamma\gamma$ using the following set of cuts, 
 \begin{eqnarray}
 p_{T}^{\gamma} > 20 \,{\rm{GeV}} \;, \quad |\eta_{\gamma}| < 2.5 \;, \quad R_{\gamma\gamma} > 0.4 \;, \nonumber\\*
 p_{T}^{\ell_h} > 25  \,{\rm{GeV}} \;, \quad p_{T}^{\ell_s} > 15 \,{\rm{GeV}} \;, \quad |\eta_{\ell}| < 2.5 \;, \nonumber\\*
 m_{\ell\ell} > 12  \,{\rm{GeV}} \;, \quad R_{\ell \gamma} > 0.7 \;, \quad E_{T}^{\mathrm{miss}} > 25 \,{\rm{GeV}} \;,
 \label{eq:zgamgamcuts}
 \end{eqnarray}  
  where $p_{T}^{\ell_h,s}$ are the transverse momenta of the harder ($h$) and softer ($s$) leptons of the pair.   These cuts are typical of those used in
 selection criteria at the LHC. For simplicity we maintain the same photon isolation requirements as in the previous section. For the common renormalization,
 factorization and fragmentation scale we choose $\mu=m_{\ell\ell\gamma\gamma}$.  Cross sections as a function of $\sqrt{s}$ are shown in
 Fig.~\ref{fig:gamgam_xs}, together with the usual scale variation.
 The total uncertainty on the cross section, including variations of both PDF and fragmentation
 sets, is rather small. For instance, the cross section for $\ell^+\ell^-\gamma\gamma$ at $8$~TeV is,
\begin{eqnarray*}
\sigma^{NLO}(\ell^+\ell^- \gamma \gamma) = 2.29 \,\rm{fb} ^{+2.3 \% } _{-1.6 \%}~\mbox{(scale)} 
\pm 4.1 \%~\mbox{(PDF)}  \pm 0.5\%~\mbox{(frag)} \;.
\end{eqnarray*}  
 The $K$-factor across the range of operating energies considered is approximately $1.5$ for the $\ell^+\ell^-\gamma\gamma$ final state and $1.65$ for
 $\nu\bar\nu\gamma\gamma$. 
 The large $K$-factors for these processes are reminiscent of that obtained for $Z\gamma$~\cite{Campbell:2011bn}. In both cases 
 the underlying Born process is $q\bar q$-initiated and therefore the real corrections are significant, due to the large PDF enhancement from gluons in the initial
 state.

\begin{figure}
\begin{center} 
\includegraphics[scale=1.4,angle=0]{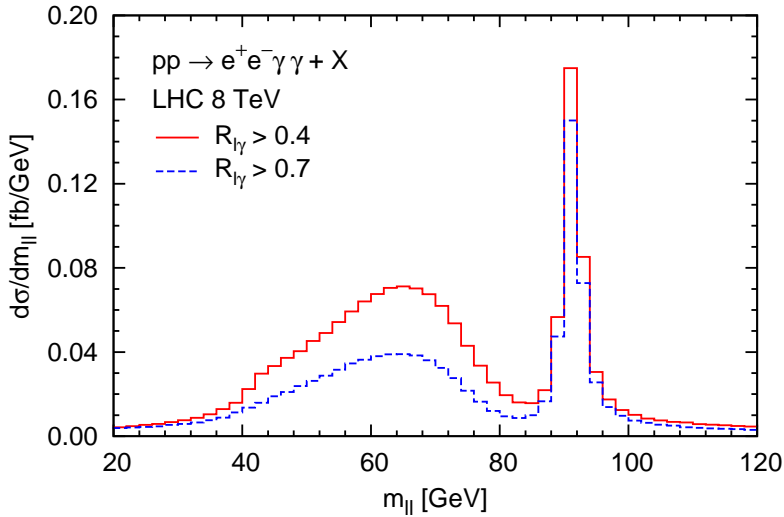} 
\caption{NLO predictions for the invariant mass of the electron-positron pair in $e^+e^-\gamma\gamma$ events at 8 TeV.
The dashed (blue) histogram shows the prediction using our usual electron-photon separation cut, $R_{\ell\gamma}>0.7$ while
the solid (red) prediction is for the looser cut,  $R_{\ell\gamma}>0.4$.}
\label{fig:mll_zaa}
\end{center}
\end{figure}

 We note that the cut  we have used on $m_{\ell\ell}$ is quite low, i.e we allow the lepton pair to be a long  way from the $Z$ pole. This benefits the analysis
 since there are a large number of $\ell^+\ell^-\gamma\gamma$ events that contain at least one photon radiated from the leptons, thereby reducing $m_{\ell\ell}$.
 In Fig.~\ref{fig:mll_zaa} we present the invariant mass distribution of the two charged leptons using two values of the lepton-photon separation cut.
 We compare our usual choice presented in Eq.~(\ref{eq:zgamgamcuts}), i.e. $R_{\ell\gamma} > 0.7$, with the looser requirement $R_{\ell\gamma} > 0.4$.  
 From this figure it is clear that the invariant mass of
 the leptons is often away from the $Z$ window. As expected, relaxing the isolation requirement dramatically  enhances this region of phase space since the
 radiation peaks in the collinear region. These results suggest that the lepton-photon isolation should be as loose as is experimentally feasible in order
 to enhance the rate. 

\section{Conclusions}
\label{sec:conc}

In this paper we have calculated NLO corrections to the production of $Z\gamma$+jet and $Z\gamma\gamma$ at hadron colliders. 
We have included the full decays of the $Z$ boson to leptons including, where appropriate, the radiation of photons from 
the $Z$ decay products. We include fragmentation contributions in order that photons can be isolated in a manner that is
analogous to the current experimental procedure.
Our results have been included in the Monte Carlo program {MCFM} 
that is publicly available. 

The results presented in this paper were obtained using analytic expressions for helicity amplitudes. In order to build these amplitudes 
the relevant Feynman diagrams were separated into gauge invariant subsets based on the identity of the fermion radiating the photon.
This allowed an efficient
recycling of earlier results for $Zj$ and $Zjj$ amplitudes into the necessary ingredients for the calculations at hand. 
The amplitudes in which a photon is radiated from a quark line are related to colour suppressed pieces of the $q\overline{q}Z gg$ amplitudes originally calculated in Refs.~\cite{bdk1,Nagy:1998bb}. The
remaining amplitudes, corresponding to photon radiation from the leptons, are presented here for the first  time. They were obtained by extracting suitable QCD currents from Ref.~\cite{bdk1} and then
contracting them with the current for a $Z$ boson decaying to leptons and one or two photons.

We have studied the phenomenology associated with the production of a $Z$-boson in association with a photon and zero or one jets at NLO.
The ATLAS collaboration has recently presented results for $\ell\ell\gamma$ cross sections~\cite{Aad:2012mr}, separating their results
into bins classified by the number of jets present. Theoretical predictions for cross sections with a specified number of jets are
subject to large uncertainties that can be underestimated by traditional estimates of the scale uncertainty~\cite{Stewart:2011cf}.
We have shown that using the method of Ref.~\cite{Stewart:2011cf} indeed provides a more reasonable estimate of the theoretical uncertainty.
We also consider both PDF uncertainties, by using the 68\% confidence limit of CT10~\cite{Lai:2010vv}, and fragmentation uncertainties,
by comparing  two independent fragmentation function calculations.
 We are thus able to provide theoretical predictions for the binned cross sections with uncertainties estimated using the best
 available information.
We also studied the photon $p_T$ spectrum in the various bins and presented ratios of $p_T$ spectra associated with charged and neutral
leptonic decays of the $Z$-boson. 

Finally, we considered the much rarer $Z\gamma\gamma$ process. 
Since the rate for this process is quite small it has so far not been observed at a hadron collider. In anticipation
of a future measurement after the conclusion of the $8$~TeV LHC data-taking, we have provided NLO cross sections and distributions
for this process at that operating energy.

\section*{Acknowledgements} 

We thank Keith Ellis and Al Goshaw for useful discussions. 
HBH is supported by the U.S. Department of Energy under grant DE-FG02-97ER41022 and
the Fermilab Fellowship in Theoretical Physics.
Fermilab is operated by Fermi Research Alliance, LLC under
Contract No. DE-AC02-07CH11359 with the United States Department of Energy.

\appendix

\section{Amplitudes for $Z\gamma$+jet production} 
\label{app:tree}

In this appendix we collect expressions for all the amplitudes required in the
calculation of the $Z\gamma$+jet process.

\subsection{Tree amplitudes} 


The decomposition of the tree level amplitude in terms of $q$- and $\ell$-type 
sub-amplitudes is given in Eq.~(\ref{eq:zajcolor1}).
We begin by presenting the $q$-type tree-level amplitudes. As described in Sec.~\ref{sec:zajcalc}, the expressions can be
obtained from Ref.~\cite{bdk1} by symmetrizing over the two colour orderings of the gluons in the amplitudes for
$0 \to q \qb g g \bar\ell \ell$. Explicitly we find,
\begin{eqnarray}
\aa_{q}^{(0)}(1_q^+, 2_{\qb}^-, 3_{g}^+, 4_{\gamma}^+, 5_{\bar{\ell}}^-, 6_{\ell}^+) 
&=&-i\frac{\spa1.2\spa2.5^2}{\spa 1.3\spa1.4\spa2.3\spa2.4\spa5.6} \;,  \nonumber \\
\aa_{q}^{(0)}(1_q^+, 2_{\qb}^-, 3_{g}^+, 4_{\gamma}^-, 5_{\bar{\ell}}^-, 6_{\ell}^+) 
& = & -i \left(
   \frac{\spa2.5 \spb1.2 \spb1.3 \spab5.(2+4).3}{\spb1.4 \spb2.4 \spa5.6 t_{134} t_{234}} 
+  \frac{\spa1.2 \spa2.4 \spb1.6 \spab4.(1+3).6}{\spa1.3 \spa2.3 \spb5.6 t_{134} t_{234}} \right. \nonumber \\
& & \quad \, \left. - \frac{\spab4.(1+3).6 \spab5.(2+4).3}{\spb2.4 \spa1.3 t_{134} t_{234}}
- \frac{\spa2.4 \spa2.5 \spb1.3 \spb1.6}{\spb1.4 \spa2.3 t_{134} t_{234}} \right) \;. 
\label{eq:Aqtreedef}
\end{eqnarray}
The tree-level $\ell$-type amplitudes, obtained using the procedure described in  Sec.~\ref{sec:zajcalc}, are given by, 
\begin{eqnarray}
\aa_{\ell}^{(0)}(1_q^+, 2_{\qb}^-, 3_{g}^+, 4_{\gamma}^+, 5_{\bar{\ell}}^-, 6_{\ell}^+) & = & 
i \frac{\spa2.5^2}{\spa1.3 \spa2.3 \spa4.5 \spa4.6} \;,\nonumber\\*
\aa_{\ell}^{(0)}(1_q^+, 2_{\qb}^-, 3_{g}^+, 4_{\gamma}^-, 5_{\bar{\ell}}^-, 6_{\ell}^+) & = & 
i \frac{\spab2.(1+3).6 ^2 }{ t_{456}  \spa1.3 \spa2.3 \spb4.5 \spb4.6} \;.
\label{eq:ampl1lc}
\end{eqnarray}
\subsection{Virtual amplitudes} 

We again begin with the decomposition of the $0 \rightarrow q \qb g \gamma \bar{\ell} \ell$ one-loop amplitude 
into sub-amplitudes,
\begin{eqnarray}
&& A^{(1)}(1_q, 2_{\qb}, 3_{g}, 4_{\gamma}, 5_{\bar{\ell}}, 6_{\ell}) = 2 \sqrt{2} e^3 g_s^3\cg T^{a_3}_{i_1 i_2}   \nonumber \\*
& &\times \bigg\lbrace \hspace{0.3cm} Q_q \left(-Q_q + v_{L,R}^{q}v_{L,R}^{\ell} \prop{Z}(s_{56}) \right) 
\aa^{(1)}_{q;1}(1_q, 2_{\qb}, 3_{g}, 4_{\gamma}, 5_{\bar{\ell}}, 6_{\ell}) \nonumber \\*
&& \hspace{0.6cm}  + Q_q \sum_{i=1}^{\nlf}  \left(-Q_i + \frac{1}{2} v^{\ell}_{L,R}(v^i_{L}+v^i_R) \prop{Z}(s_{56}) \right)
\aa^{(1)}_{q;2}(1_q, 2_{\qb}, 3_{g}, 4_{\gamma}, 5_{\bar{\ell}}, 6_{\ell}) \nonumber \\*
&&  \hspace{0.6cm} + Q_q \frac{v^{\ell}_{L,R}}{2 s_W c_W} \prop{Z}(s_{56}) 
\aa^{(1)}_{q;3}(1_q, 2_{\qb}, 3_{g}, 4_{\gamma}, 5_{\bar{\ell}}, 6_{\ell}) \nonumber  \\*
&& \hspace{0.6cm} + Q_{\ell} \left(-Q_q + v_{L,R}^{q}v_{L,R}^{\ell} \prop{Z}(t_{456}) \right) 
\aa^{(1)}_{\ell;1}(1_q, 2_{\qb}, 3_{g}, 4_{\gamma}, 5_{\bar{\ell}}, 6_{\ell}) \nonumber \\*
&&  \hspace{0.6cm} + Q_{\ell} \frac{v^{\ell}_{L,R}}{2 s_W c_W} \prop{Z}(t_{456}) 
\aa^{(1)}_{\ell;2}(1_q, 2_{\qb}, 3_{g}, 4_{\gamma}, 5_{\bar{\ell}}, 6_{\ell}) \bigg\rbrace, 
\label{eq:zajcolor2}
\end{eqnarray}
where the usual one-loop prefactor $c_{\Gamma}$ is defined as,
\begin{equation}
c_{\Gamma}=\frac{1}{(4 \pi)^{2-\eps}} 
\frac{\Gamma(1+\eps) \Gamma^2(1-\eps)}{\Gamma(1-2\eps)} . \\
\end{equation}
The $q$-type contributions are,
\begin{eqnarray}
\aa^{(1)}_{q;1}(1_q, 2_{\qb}, 3_{g}, 4_{\gamma}, 5_{\bar{\ell}}, 6_{\ell}) &=& 
- N_c A_6(1_q,3,2_{\qb},4) 
- \frac{1}{N_c} \bigg[ A_6(1_q,2_{\qb},3,4) + A_6(1_q,2_{\qb},4,3) \bigg] \;,\nonumber  \\*
\aa^{(1)}_{q;2}(1_q, 2_{\qb}, 3_{g}, 4_{\gamma}, 5_{\bar{\ell}}, 6_{\ell}) &=& 
A_{6;4}^{\vv}(1_q,2_{\qb},3,4) + A_{6;4}^{\vv}(1_q,2_{\qb},4,3) \;,\nonumber  \\*
\aa^{(1)}_{q;3}(1_q, 2_{\qb}, 3_{g}, 4_{\gamma}, 5_{\bar{\ell}}, 6_{\ell}) &=& 
A_{6;4}^{\ax}(1_q,2_{\qb},3,4) + A_{6;4}^{\ax}(1_q,2_{\qb},4,3)    \;,
\end{eqnarray} 
where below we list references for the primitive amplitudes from the Ref~\cite{bdk1},
\begin{center}
\begin{tabular}{rl}
$A_6(1_q,2,3_{\qb},4)$           : &Sec.~9, Eqs.~(9.2)-(9.14), \\
$A_6(1_q,2_{\qb},3,4)$           : &Sec.~10, Eqs.~(10.1)-(10.21), \\
$A_{6;4}^{\vv}(1_q,2_{\qb},3,4)$ : &Eq.~(2.13), \\
$A_{6;4}^{\ax}(1_q,2_{\qb},3,4)$ : &Eq.~(2.13). \\
\end{tabular}
\end{center}
The $\ell$-type contributions in Eq.~(\ref{eq:zajcolor2}) are,
\begin{eqnarray}
\aa^{(1)}_{\ell;1}(1_q, 2_{\qb}, 3_{g}, 4_{\gamma}, 5_{\bar{\ell}}, 6_{\ell}) &=& 
  N_c \aa^{\lc}_{\ell}(1_q, 2_{\qb}, 3_{g}, 4_{\gamma}, 5_{\bar{\ell}}, 6_{\ell} ) 
+ \frac{1}{N_c} \aa^{\slc}_{\ell}(1_q, 2_{\qb}, 3_{g}, 4_{\gamma}, 5_{\bar{\ell}}, 6_{\ell} ) \nonumber  \\*
\aa^{(1)}_{\ell;2}(1_q, 2_{\qb}, 3_{g}, 4_{\gamma}, 5_{\bar{\ell}}, 6_{\ell}) &=& 
  \aa^{\fl}_{\ell}(1_q, 2_{\qb}, 3_{g}, 4_{\gamma}, 5_{\bar{\ell}}, 6_{\ell} ) 
\label{eq:zajamp2}
\end{eqnarray}
where the sub-amplitudes have been divided into three contributions: 
leading color (lc), subleading color (sl) and fermion-loop (fl). 
We now present analytic formulae for the three contributions appearing in Eqs.~(\ref{eq:zajamp2}). We again follow the 
notation of Ref.~\cite{bdk1} and decompose our one-loop amplitude into divergent ($V$) and finite ($F$) pieces,
\begin{equation}
\aa_\ell =  V_\ell \aa_\ell^{(0)} + i F_\ell \;.
\end{equation}
We will present results for the two independent helicity combinations, $(1_q^+, 2_{\qb}^-, 3_{g}^+, 4_{\gamma}^+, 5_{\bar{\ell}}^-, 6_{\ell}^+)$
and $(1_q^+, 2_{\qb}^-, 3_{g}^+, 4_{\gamma}^-, 5_{\bar{\ell}}^-, 6_{\ell}^+)$.
The other two gluon and photon helicity combinations,
 for the same quark and lepton helicities, are obtained from the following relations,
\begin{eqnarray}
\aa_\ell(1_q^+,2_{\qb}^-,3_g^-,4_\gamma^+,5_{\bar{\ell}}^-,6_{\ell}^+) &=&
\bigg[ \aa_{\ell}(2_{\qb}^+,1_q^-,3_g^+,4_\gamma^-,6_{\ell}^-,5_{\bar{\ell}}^+)  \bigg]_{\spa i.j \leftrightarrow \spb i.j} \nonumber \\
\aa_\ell(1_q^+,2_{\qb}^-,3_g^-,4_\gamma^-,5_{\bar{\ell}}^-,6_{\ell}^+) &=&
\bigg[ \aa_{\ell}(2_{\qb}^+,1_q^-,3_g^+,4_\gamma^+,6_{\ell}^-,5_{\bar{\ell}}^+)  \bigg]_{\spa i.j \leftrightarrow \spb i.j} \nonumber 
\end{eqnarray}
For the case where the helicity of the lepton pair is flipped, 
the amplitudes are obtained by performing a $5 \leftrightarrow 6$ exchange with
an extra minus sign due to the $Z/\gamma^* \rightarrow \ell^+ \ell^- \gamma$ current,
\begin{eqnarray}
\aa_\ell(1_q,2_{\qb},3_g,4_\gamma,5_{\bar{\ell}}^+,6_{\ell}^-) &=&
\left.- \aa_\ell(1_q,2_{\qb},3_g,4_\gamma,5_{\bar{\ell}}^-,6_{\ell}^+)\right|_{5 \leftrightarrow 6} \nonumber .
\end{eqnarray}
Similarly, the amplitudes for the remaining helicity combinations can be obtained via,
\begin{eqnarray}
\aa_\ell(1_q^-,2_{\qb}^+,3_g,4_\gamma,5_{\bar{\ell}},6_{\ell}) &=&
\left. - \aa_\ell(1_q^+,2_{\qb}^-,3_g,4_\gamma,5_{\bar{\ell}},6_{\ell})\right|_{1 \leftrightarrow 2} \nonumber .
\end{eqnarray}
All helicities share a common divergent factor,
\begin{equation}
V_{\ell}^{\lc} = -\frac{1}{\eps^2} \left[  \left( \frac{\musq}{-s_{13}} \right)^{\eps} 
                                      + \left( \frac{\musq}{-s_{23}} \right)^{\eps} \right]
              -\frac{3}{2 \eps} \left( \frac{\musq}{-t_{456}} \right)^{\eps} -3 \;.
\label{eq:ampl2lc}
\end{equation}
For the helicity configuration $(1_q^+, 2_{\qb}^-, 3_{g}^+, 4_{\gamma}^+, 5_{\bar{\ell}}^-, 6_{\ell}^+)$
the finite remainder is,
\begin{eqnarray}
F_{\ell}^{\lc} & = & 
   - \frac{\spa2.5 ^2}{\spa1.3 \spa2.3 \spa4.5 \spa4.6}  
     \Ls_{-1}\left( \frac{ -s_{13}}{-t_{456}}, \frac{-s_{23}}{-t_{456}}  \right) 
   - \frac{ \spa1.2 \spa2.5 \spab5.(4+6).1 }{\spa1.3 \spa2.3 \spa4.5 \spa4.6 } 
     \frac{\lzero{\frac{-s_{23}}{-t_{456}}}}{t_{456}} \nonumber \\*
&& + \frac{ \spa1.2 ^2 \spab5.(2+3).1 \spab5.(4+6).1 }{ 2 \spa1.3 \spa2.3 \spa4.5 \spa4.6 }  
     \frac{\lone{\frac{-s_{23}}{-t_{456}}}}{t_{456}^2} ,
\label{eq:ampl3lc}
\end{eqnarray}
and for $(1_q^+, 2_{\qb}^-, 3_{g}^+, 4_{\gamma}^-, 5_{\bar{\ell}}^-, 6_{\ell}^+)$,
\begin{eqnarray}
F_{\ell}^{\lc} & = & 
     \frac{\spab2.(1+3).6 \spab2.(4+5).6}{ t_{456} \spa1.3 \spa2.3 \spb4.5 \spb4.6 }
     \Ls_{-1}\left( \frac{ -s_{13}}{-t_{456}}, \frac{-s_{23}}{-t_{456}} \right)  
   - \frac{ \spab2.1.6 \spab2.(4+5).6) }{ \spa1.3 \spa2.3 \spb4.5 \spb4.6 } 
     \frac{\lzero{\frac{-s_{23}}{-t_{456}}}}{t_{456}} \nonumber  \\*
&& - \frac{ \spb1.2 ^2 \spb1.6 ^2 }{ 2 \spa1.3 \spa2.3 \spb4.5 \spb4.6 } 
     \frac{\lone{\frac{-s_{23}}{-t_{456}}}}{t_{456}^2} .
\label{eq:ampl4lc}
\end{eqnarray}
Explicit formulae for the basis integrals appearing in this formulae can be found in Appendix II of Ref.~\cite{bdk1}. 

Next we consider the sub-leading colour contribution to  
$\aa_{\ell}(1_q^+, 2_{\qb}^-, 3_{g}^+, 4_{\gamma}^+, 5_{\bar{\ell}}^-, 6_{\ell}^+)$.
Using the same notation as before the divergent pieces are given by,
\begin{equation}
V_{\ell}^{\slc} = \frac{1}{\eps^2} \left( \frac{\musq}{-s_{12}} \right)^{\eps} 
               +\frac{3}{2 \eps} \left( \frac{\musq}{-t_{456}} \right)^{\eps} +\frac{7}{2} \;.
\label{eq:ampl2slc}
\end{equation}
The finite part for the $(1_q^+, 2_{\qb}^-, 3_{g}^+, 4_{\gamma}^+, 5_{\bar{\ell}}^-, 6_{\ell}^+)$ configuration is.
\begin{eqnarray}
F_{\ell}^{\slc} & = & 
  \frac{\spa2.5 ^2}{\spa1.3 \spa2.3 \spa4.5 \spa4.6}  
  \Ls_{-1}\left( \frac{ -s_{12}}{-t_{456}}, \frac{-s_{13}}{-t_{456}}  \right)
+ \frac{\spa1.2 ^2 \spa3.5 ^2}{\spa1.3 ^3 \spa2.3 \spa4.5 \spa4.6}  
  \Ls_{-1}\left( \frac{ -s_{12}}{-t_{456}}, \frac{-s_{23}}{-t_{456}}  \right) \nonumber \\*
&& - \frac{ 2 s_{13} \spa1.5 \spa2.5 - \spab2.3.1 \spa1.5 ^2  }{\spa1.3 ^2 \spa4.5 \spa4.6 } 
     \frac{\lzero{\frac{-t_{456}}{-s_{23}}}}{s_{23}} 
   - \frac{ \spb1.3 ^2 \spa2.3 \spa1.5 ^2 }{ 2 \spa1.3 \spa4.5 \spa4.6 }  
     \frac{\lone{\frac{-t_{456}}{-s_{23}}}}{s_{23}^2} \nonumber \\*
&& - \frac{ \spab2.1.3 \spa1.5 \spa3.5  }{\spa1.3 ^2 \spa4.5 \spa4.6 } 
     \frac{\lzero{\frac{-t_{456}}{-s_{12}}}}{s_{12}}  
   + \frac{ \spab2.1.3 \spa5.3 \spab5.(4+6).3  }{ \spa1.3 \spa4.5 \spa4.6 }   
     \frac{\lone{\frac{-t_{456}}{-s_{12}}}}{s_{12}^2} \\*
&& - \frac{ \spab5.(1+2).3 \left( \spb1.3 \spab5.(4+6).2 + \spb2.3 \spab5.(4+6).1 \right) }
     {2 t_{456} \spb1.2 \spb2.3 \spa1.3 \spa4.5 \spa4.6  } \nonumber
\label{eq:ampl3slc}
\end{eqnarray}
while the corresponding contribution for $(1_q^+, 2_{\qb}^-, 3_{g}^+, 4_{\gamma}^-, 5_{\bar{\ell}}^-, 6_{\ell}^+)$  is,
\begin{eqnarray}
F_{\ell}^{\slc} & = & 
- \frac{\spab2.(1+3).6 \spab2.(4+5).6}{ t_{456} \spa1.3 \spa2.3 \spb4.5 \spb4.6 }
  \Ls_{-1}\left( \frac{ -s_{12}}{-t_{456}}, \frac{-s_{13}}{-t_{456}} \right) \nonumber \\*
&& - \frac{\spa1.2 ^2 \spab3.(1+2).6 \spab3.(4+5).6}{t_{456} \spa1.3^3 \spa2.3 \spb4.5 \spb4.6} 
     \Ls_{-1}\left( \frac{ -s_{12}}{-t_{456}}, \frac{-s_{23}}{-t_{456}}  \right) \nonumber \\*
&& + \frac{ \spab1.(2+3).6 (2 s_{13} \spab2.(4+5).6 - \spab2.3.1 \spab1.(4+5).6) }{t_{456} \spa1.3 ^2 \spb4.5 \spb4.6 } 
     \frac{\lzero{\frac{-t_{456}}{-s_{23}}}}{s_{23}}  \\*
&& + \frac{ \spb1.3 ^2 \spa2.3 \spab1.(2+3).6 \spab1.(4+5).6 }{ 2 t_{456} \spa1.3 \spb4.5 \spb4.6 } 
     \frac{\lone{\frac{-t_{456}}{-s_{23}}}}{s_{23}^2} \nonumber \\*
&& + \frac{ \spab2.1.3 \spab3.(1+2).6 \spab1.(4+5).6 }{ t_{456} \spa1.3 ^2 \spb4.5 \spb4.6 } 
     \frac{\lzero{\frac{-t_{456}}{-s_{12}}}}{s_{12}}  
   + \frac{ \spab2.1.3 \spa6.3 \spab3.(4+5).6  }{ \spa1.3 \spb4.5 \spb4.6 }
     \frac{\lone{\frac{-t_{456}}{-s_{12}}}}{s_{12}^2} \nonumber \\*
&& - \frac{ \spb3.6 ( \spb1.3 \spb6.2 + \spb2.3 \spb6.1) }{2 \spb1.2 \spb2.3 \spa1.3 \spb4.5 \spb4.6  }. \nonumber
\label{eq:ampl4slc}
\end{eqnarray}
Lastly we consider the fermion loop contribution, which does not contain a divergent piece ($V_\ell^{\fl}= 0$).
The finite part for the
$(1_q^+, 2_{\qb}^-, 3_{g}^+, 4_{\gamma}^+, 5_{\bar{\ell}}^-, 6_{\ell}^+)$ configuration is
\begin{eqnarray}
F_{\ell}^{\fl} & = & 
 \frac{ \spb1.3 \spa2.5 \spab5.(4+6).3 }{ \spa4.5 \spa4.6 }  
 \left[ \frac{\lone{\frac{-s_{12}}{-t_{456}}}}{t_{456}^2} - \frac{1}{12 t_{456} m_t^2} \right].
\label{eq:ampl1fl}
\end{eqnarray}
and the result for the the other helicity configuration $(1_q^+, 2_{\qb}^-, 3_{g}^+, 4_{\gamma}^-, 5_{\bar{\ell}}^-, 6_{\ell}^+)$ is,
\begin{eqnarray}
F_{\ell}^{\fl} & = & 
 \frac{ \spb1.3 \spb3.6 \spab2.(4+5).6 }{ \spb4.5 \spb4.6 }  
 \left[ \frac{\lone{\frac{-s_{12}}{-t_{456}}}}{t_{456}^2} - \frac{1}{12 t_{456} m_t^2} \right].
\label{eq:ampl2fl}
\end{eqnarray}
Here the top quark loop contribution is expanded in powers of $1/m_t^2$ and the terms of order $1/m_t^4$ are dropped.

\subsection{Real emission amplitudes}
\label{subsec:zajreal}

Next we consider the real corrections to the $Z \gamma$+jet process. We use the same 
notation as in the previous section to designate $q$- and $\ell$- type diagrams.  

\subsubsection{Amplitudes for $0 \to q \qb g g \gamma \bar{\ell} \ell$}
We begin by considering the amplitudes containing only one quark line, i.e. the process $0 \rightarrow q \qb g g \gamma \bar{\ell} \ell$.  
The decomposition of this amplitude is given by,
\begin{eqnarray}
&& A^{(0)}(1_q, 2_\qb, 3_{g}, 4_{g}, 5_{\gamma}, 6_{\bar{\ell}}, 7_{\ell}) = 2 \sqrt{2} e^3 g_s^2 
\;\; \sum_{ \lbrace 3,4 \rbrace}  \left(T^{a_3} T^{ a_4} \right)_{i_1 i_2} \nonumber \\*
&&\times \bigg\lbrace \hspace{0.3cm} Q_q \left(-Q_q + v_{L,R}^{q}v_{L,R}^{\ell} \prop{Z}(s_{67}) \right) 
  \aa^{(0)}_{q}\left(1_q, 2_{\qb}, 3_{g}, 4_{g}, 5_{\gamma}, 6_{\bar{\ell}}, 7_{\ell}\right) \nonumber \\*
&&\hspace{0.6cm} + Q_{\ell} \left(-Q_q + v_{L,R}^{q}v_{L,R}^{\ell} \prop{Z}(t_{567}) \right)
  \aa^{(0)}_{\ell}\left(1_q, 2_{\qb}, 3_{g}, 4_{g}, 5_{\gamma}, 6_{\bar{\ell}}, 7_{\ell}\right) \bigg\rbrace \;.
\label{eq:zajjcolor1}
\end{eqnarray}
All that has changed with respect to the tree-level decomposition is the presence of the additional gluon, which manifests itself 
in the form of the amplitudes and the colour structure.

The helicity amplitudes for diagrams in which the photon is emitted from the quark line are obtained from the amplitudes
for the $e^+ e^- \rightarrow q \qb g g g$ process presented in Ref.~\cite{Nagy:1998bb}, where one gluon is replaced by a photon.
\begin{eqnarray}
\aa^{(0)}_{q}\left(1_q,2_{\qb},3_{g},4_{g},5_{\gamma},6_{\bar{\ell}},7_{\ell}\right) &=& \frac{i}{s_{67}} 
\bigg[ A(1_q,3_g,4_g,5_g,2_{\qb}) + A(1_q,3_g,5_g,4_g,2_{\qb}) \nonumber \\*
&& \hspace{0.8cm} + A(1_q,5_g,3_g,4_g,2_{\qb})  \bigg]
\label{eq:zajjamp1}
\end{eqnarray}
where on the right-hand side, the 6, 7 labels for lepton pair have been suppressed.
The analytic expressions for  $A(1_q,3_g,4_g,5_g,2_{\qb})$ are presented 
in Appendix A, Eqs.~(A43)-(A49) of Ref.~\cite{Nagy:1998bb}.
The helicity amplitudes for diagrams where the photon is emitted from the lepton line are given by
\begin{eqnarray} 
&& \aa^{(0)}_{\ell}(1^+_q,2^-_{\qb} , 3^+_{g}, 4^+_{g}, 5^+_{\gamma}, 6^-_{\bar{\ell}}, 7^+_{\ell}) =
i \frac{\spa2.6 ^2}{\spa1.3 \spa3.4 \spa4.2 \spa5.6 \spa5.7} \;,\label{eq:zajjamp2} \\
&& \aa^{(0)}_{\ell}(1^+_q,2^-_{\qb} , 3^+_{g}, 4^+_{g},5^-_{\gamma} , 6^-_{\bar{\ell}}, 7^+_{\ell}) =
i \frac{\spab2.(5+6).7 ^2}{t_{567} \spa1.3 \spa3.4 \spa4.2 \spb5.6 \spb5.7} \;, \\
&& \aa^{(0)}_{\ell}(1^+_q,2^-_{\qb} , 3^-_{g}, 4^-_{g},5^+_{\gamma} , 6^-_{\bar{\ell}}, 7^+_{\ell}) =
-i \frac{\spab6.(5+7).1 ^2}{t_{567} \spb1.3 \spb3.4 \spb4.2 \spa5.6 \spa5.7} \;, \\
&& \aa^{(0)}_{\ell}(1^+_q,2^-_{\qb} , 3^-_{g}, 4^-_{g},5^-_{\gamma} , 6^-_{\bar{\ell}}, 7^+_{\ell}) =
-i \frac{\spb1.7 ^2}{\spb1.3 \spb3.4 \spb4.2 \spb5.6 \spb5.7} \;, \\
&& \aa^{(0)}_{\ell}(1^+_q,2^-_{\qb} , 3^+_{g}, 4^-_{g},5^+_{\gamma} , 6^-_{\bar{\ell}}, 7^+_{\ell}) =
\frac{i}{s_{34} t_{567} \spa5.6 \spa5.7} 
\left\lbrace \frac{\spab4.1.3 \spa2.6 \spaa4.(1+3)(5+7).6}{\spa1.3 t_{134}} \right. \nonumber \\
& & \left. -\frac{\spab4.2.3 \spab6.(4+2).3 \spab6.(5+7).1}{\spb4.2 t_{234}}
-\frac{\spaa4.(1+3)(5+7).6 \spab6.(4+2).3}{\spa1.3 \spb4.2} \right\rbrace \;, \\
&& \aa^{(0)}_{\ell}(1^+_q,2^-_{\qb} , 3^+_{g}, 4^-_{g},5^-_{\gamma} , 6^-_{\bar{\ell}}, 7^+_{\ell}) =
\frac{i}{s_{34} t_{567} \spb5.6 \spb5.7} 
\left\lbrace \frac{\spab4.1.3 \spab4.(1+3).7 \spab2.(5+6).7}{\spa1.3 t_{134}} \right. \nonumber \\
& & \left. -\frac{\spab4.2.3 \spbb3.(4+2)(5+6).7 \spb1.7 }{\spb4.2 t_{234}}
-\frac{ \spbb3.(4+2)(5+6).7 \spab4.(1+3).7}{\spa1.3 \spb4.2} \right\rbrace \;, \\
&& \aa^{(0)}_{\ell}(1^+_q,2^-_{\qb} , 3^-_{g}, 4^+_{g},5^+_{\gamma} , 6^-_{\bar{\ell}}, 7^+_{\ell}) =
\frac{i}{s_{34} t_{567} \spa5.6 \spa5.7} 
\left\lbrace  -\frac{\spb1.4 ^2 \spa2.6\spaa3.(1+4)(5+7).6}{\spb1.3 t_{134}} \right. \nonumber \\
& & \left. + \frac{\spa3.2 ^2 \spab6.(3+2).4 \spab6.(5+7).1}{\spa4.2 t_{234}}
+ \frac{\spb1.4 \spa3.2 \spa2.6 \spab6.(5+7).1 }{\spb1.3 \spa4.2} \right\rbrace \;, \\
&& \aa^{(0)}_{\ell}(1^+_q,2^-_{\qb} , 3^-_{g}, 4^+_{g},5^-_{\gamma} , 6^-_{\bar{\ell}}, 7^+_{\ell}) =
\frac{i}{s_{34} t_{567} \spb5.6 \spb5.7} 
\left\lbrace -\frac{\spb1.4 ^2 \spab3.(1+4).7 \spab2.(5+6).7}{\spb3.4 t_{134}} \right. \nonumber \\
& & \left. + \frac{\spa3.2 ^2 \spb1.7 \spbb4.(3+2)(5+6).7 }{\spa4.2 t_{234}}
- \frac{\spb1.4 \spa3.2 \spb7.1 \spab2.(5+6).7}{\spb1.3 \spa4.2} \right\rbrace \;. \label{eq:zajjamp3}
\end{eqnarray}

\subsubsection{Amplitudes for $0 \to q \qb Q \bar Q \gamma  \bar{\ell} \ell$}
Next we consider the processes that contain two quark lines. The decomposition of the amplitude is,
\begin{eqnarray}
&& A^{(0)}(1_q, 2_{\qb}, 3_Q, 4_{\bar Q},  5_{\gamma}, 6_{\bar{\ell}}, 7_{\ell}) = 
2 \sqrt{2} e^3 g_s \; T^{b}_{i_1 i_2} T^{b}_{i_3 i_4} \delta_{q \qb} \delta_{Q \bar Q}  \nonumber \\*
&&\times \bigg[\left(-Q_q + v_{L,R}^{q}v_{L,R}^{\ell} \prop{Z}(s_{67}) \right)
\aa^{(0)}_{q}(1_q,  2_{\qb}, 3_Q, 4_{\bar Q}, 5_{\gamma}, 6_{\bar{\ell}}, 7_{\ell}) \nonumber \\
&& \hspace{0.6cm} +Q_{\ell}\left(-Q_q + v_{L,R}^{q}v_{L,R}^{\ell} \prop{Z}(t_{567}) \right)
\aa^{(0)}_{\ell}(1_q,  2_{\qb}, 3_Q, 4_{\bar Q}, 5_{\gamma}, 6_{\bar{\ell}}, 7_{\ell}) \\
&& \hspace{0.6cm} +\left(-Q_Q + v_{L,R}^{Q}v_{L,R}^{\ell} \prop{Z}(s_{67}) \right)
\aa^{(0)}_{q}(3_Q,  4_{\bar Q}, 1_q, 2_{\bar q}, 5_{\gamma}, 6_{\bar{\ell}}, 7_{\ell})   \nonumber \\
&& \hspace{0.6cm} +Q_{\ell}\left(-Q_Q + v_{L,R}^{Q}v_{L,R}^{\ell} \prop{Z}(t_{567}) \right)
\aa^{(0)}_{\ell}(3_Q,  4_{\bar Q}, 1_q, 2_{\bar q}, 5_{\gamma}, 6_{\bar{\ell}}, 7_{\ell})  \bigg]
 - \bigg\{ 2_\qb \leftrightarrow 4_{\bar Q} \bigg\} \;. \nonumber 
\end{eqnarray}
The four quark amplitudes are obtained from Ref.~\cite{Nagy:1998bb} in a similar fashion as described above.
The helicity amplitudes for diagrams where the photon is emitted from the quark line are given by,
\begin{eqnarray}
&&\aa^{(0)}_{q}(1_q, 2_{\bar q}, 3_Q, 4_{\bar Q}, 5_{\gamma}, 6_{\bar{\ell}}, 7_{\ell}) =
 \frac{i}{s_{67}} \bigg[\hspace{0.3cm} Q_q \left( A_1(1_q,2_{\qb},3_Q,4_{\bar Q},5_g) 
+ A_2(1_q,2_{\qb},3_Q,4_{\bar Q},5_g) \right) \nonumber \\
&&\hspace{0.8cm} + Q_Q \left( A_3(1_q,2_{\qb},3_Q,4_{\bar Q},5_g) 
+ A_4(1_q,2_{\qb},3_Q,4_{\bar Q},5_g) \right) \bigg] \;.
\end{eqnarray}
The analytic expressions for $A_i(1_q,2_{\qb},3_Q,4_{\bar Q},5_g)$ are presented 
in Appendix A, Eqs.~(A50)-(A59) of Ref.~\cite{Nagy:1998bb}.
For diagrams where the photon is emitted from the lepton line the helicity amplitudes are, 
\begin{eqnarray}
&&\aa^{(0)}_{\ell}(1^+_q, 2^-_{\bar q}, 3^+_Q, 4^-_{\bar Q}, 5^+_{\gamma}, 6^-_{\bar{\ell}}, 7^+_{\ell}) =
\frac{-i}{s_{34} t_{567} \spa5.6 \spa5.7 } 
\nonumber \\ &\times& \Biggl\{ 
   \frac{\spb1.3 \spa2.6 \spaa4.(1+3)(5+7).6}{t_{134}}
 + \frac{\spa4.2 \spab6.(2+4).3 \spab6.(5+7).1}{t_{234}} \Biggr\}  \;, \\
&& \aa^{(0)}_{\ell}(1^+_q, 2^-_{\bar q}, 3^+_Q, 4^-_{\bar Q}, 5^-_{\gamma}, 6^-_{\bar{\ell}}, 7^+_{\ell}) =
\frac{i}{s_{34} t_{567} \spb5.6 \spb5.7 } 
 \nonumber \\ &\times& \Biggl\{ 
    - \frac{\spb1.3 \spab4.(1+3).7 \spab2.(5+6).7}{t_{134}} 
    + \frac{\spa4.2 \spb7.1 \spbb3.(2+4)(5+6).7}{t_{234}} \Biggr\} \;.
\end{eqnarray}

\section{Amplitudes for $Z\gamma\gamma$ production} 
\label{app:zgamgam}

The decomposition of the tree level $0 \to q \qb \gamma \gamma \bar\ell \ell$ amplitudes in terms of $qq$-,$q\ell$- and $\ell\ell$-type 
sub-amplitudes is given in Eq.~(\ref{eq:zaacolor1}).
These sub-amplitudes can be obtained from the results of the previous section using the
following relations, 
\begin{eqnarray}
\aa^{(0)}_{qq}(1 _q, 2_{\qb}, 3_{\gamma}, 4_{\gamma}, 5_{\bar{\ell}}, 6_{\ell}) 
&=& \aa^{(0)}_{q}(1 _q, 2_{\qb}, 3_{g}, 4_{\gamma}, 5_{\bar{\ell}}, 6_{\ell}) \\
\aa^{(0)}_{q\ell}(1 _q, 2_{\qb}, 3_{\gamma}, 4_{\gamma}, 5_{\bar{\ell}}, 6_{\ell}) 
&=& \aa^{(0)}_{\ell}(1 _q, 2_{\qb}, 3_{g}, 4_{\gamma}, 5_{\bar{\ell}}, 6_{\ell}) 
\label{eq:Aqltreedef}  \\
\aa^{(0)}_{\ell\ell}(1_q, 2_{\qb}, 3_{\gamma}, 4_{\gamma}, 5_{\bar{\ell}}, 6_{\ell}) 
&=& \aa^{(0)}_{qq}(6_{\ell}, 5_{\bar{\ell}}, 3_{\gamma}, 4_{\gamma},  2_{\qb}, 1_q),
\label{eq:ampqq1}
\end{eqnarray}
The amplitudes $\aa^{(0)}_{q}$ and $\aa^{(0)}_\ell$ are defined in Eqs.(\ref{eq:Aqtreedef})
and~(\ref{eq:ampl1lc}) respectively.
Note that Eq.~(\ref{eq:Aqltreedef}) represents the tree-level amplitude for the emission of photon
4 from the lepton line.

Due to the simple colour structure, the extension of Eq.~(\ref{eq:zaacolor1})
to the one-loop case is simple,
\begin{eqnarray}
A^{(1)}(1_q,2_{\qb},3_{\gamma},4_{\gamma},5_{\bar{\ell}},6_{\ell}) &=& \cg g_s^2 \left( N_c -\frac{1}{N_c} \right) 
A^{(0)}(1_q,2_{\qb},3_{\gamma},4_{\gamma},5_{\bar{\ell}},6_{\ell}) \,\, \Bigl\{ \aa^{(0)}_{XY} \rightarrow \aa^{(1)}_{XY} \Bigr\} \;.\nonumber\\*
\end{eqnarray}
Analogous to the $Z\gamma$+jet process, the helicity amplitudes for the $qq$-type diagrams can be obtained 
from the amplitudes for the $e^+ e^- \rightarrow q \qb g g$ process presented in Ref.~\cite{bdk1}. 
In particular we only need the subleading color primitive amplitude $A_6(1_q,2_{\qb},3,4)$ 
symmetrized over the two gluons.  
The $q\ell$-type amplitudes can be obtained in a similar fashion from the subleading color contribution 
to $\aa_{\ell}(1_q^+,2_{\qb}^-,3_{g}^+,4_{\gamma}^+,5_{\bar{\ell}}^-,6_{\ell}^+)$ amplitude, which we presented in Appendix~\ref{app:tree}. 
Specifically in terms of the amplitudes presented in \cite{bdk1} and the previous section we have,
\begin{eqnarray}
\aa^{(1)}_{qq}(1 _q,2_{\qb},3_{\gamma},4_{\gamma},5_{\bar{\ell}},6_{\ell}) 
&=& A_6(1_q,2_{\qb},3,4) + A_6(1_q,2_{\qb},4,3) \;,\\
\aa^{(1)}_{q\ell}(1_q, 2_{\qb}, 3_{\gamma}, 4_{\gamma}, 5_{\bar{\ell}}, 6_{\ell}) 
&=& -A_\ell^{\slc}(1_q,2_{\qb},3_{g},4_{\gamma},5_{\bar{\ell}},6_{\ell}) \;.
\label{eq:ampqq2}
\end{eqnarray}

The $\ell\ell$-type contribution to the $0 \rightarrow  q \qb \gamma \gamma \bar{\ell} \ell$ one-loop amplitude 
simply consists of a vertex correction and is given by
\begin{eqnarray}
\aa^{(1)}_{\ell\ell}(1_q,2_{\qb},3_{\gamma},4_{\gamma},5_{\bar{\ell}},6_{\ell}) & = & \left(\frac{\musq}{-s_{12}}\right)^{\eps}
\left[-\frac{1}{\eps^2}-\frac{3}{2 \eps}-4\right] \aa^{(0)}_{\ell\ell}(1_q,2_{\qb},3_{\gamma},4_{\gamma},5_{\bar{\ell}},6_{\ell}).
\label{eq:ampll2}
\end{eqnarray}

Finally we consider the real corrections in which an additional gluon is radiated relative to the Born process. In terms of our usual 
decomposition the $0 \rightarrow q \qb g \gamma \gamma \bar{\ell} \ell$ amplitude is given by,
\begin{eqnarray}
&& A^{(0)}(1_q,2_{\qb},3_g,4_{\gamma},5_{\gamma},6_{\bar{\ell}},7_{\ell}) = 4 e^4 g_s T^{a_3}_{i_1 i_2} \nonumber \\*
&& \times \bigg[ Q_q^2 \left(-Q_q + v_{L,R}^{q}v_{L,R}^{\ell} \prop{Z}(s_{67}) \right)
\aa^{(0)}_{qq}(1_q,2_\qb,3_g,4_{\gamma},5_{\gamma},6_{\bar{\ell}},7_{\ell}) \nonumber \\*
&& \hspace{0.6cm} +Q_{\ell}Q_q \left(-Q_q + v_{L,R}^{q}v_{L,R}^{\ell} \prop{Z}(t_{467}) \right)
\aa^{(0)}_{q\ell}(1_q,2_\qb,3_g,4_{\gamma},5_{\gamma},6_{\bar{\ell}},7_{\ell}) \nonumber \\*
&& \hspace{0.6cm} +Q_{\ell}Q_q \left(-Q_q + v_{L,R}^{q}v_{L,R}^{\ell} \prop{Z}(t_{567}) \right)
\aa^{(0)}_{q\ell}(1_q,2_\qb,3_g,4_{\gamma},5_{\gamma},6_{\bar{\ell}},7_{\ell}) \nonumber \\*
&& \hspace{0.6cm}  +Q_{\ell}^2 \left(-Q_q + v_{L,R}^{q}v_{L,R}^{\ell} \prop{Z}(t_{4567}) \right)
\aa^{(0)}_{\ell\ell}(1_q,2_\qb,3_g,4_{\gamma},5_{\gamma},6_{\bar{\ell}},7_{\ell})
\bigg]. 
\label{eq:zaajcolor1}
\end{eqnarray}
The sub-amplitudes can be constructed as follows,
\begin{eqnarray}
&& \aa^{(0)}_{qq}\left(1_q,2_\qb,3_g,4_{\gamma},5_{\gamma},6_{\bar{\ell}},7_{\ell}\right) = 
  \aa^{(0)}_{q}\left(1_q,2_{\qb},3_{g},4_{g},5_{\gamma},6_{\bar{\ell}},7_{\ell}\right)
+ \aa^{(0)}_{q}\left(1_q,2_{\qb},3_{g},5_{\gamma},4_{g},6_{\bar{\ell}},7_{\ell}\right) \;, \nonumber \\*
&& \aa^{(0)}_{q\ell}\left(1_q,2_{\qb}, 3_{g}, 4_{\gamma}, 5_{\gamma}, 6_{\bar{\ell}}, 7_{\ell}\right) = 
  \aa^{(0)}_{\ell}\left(1_q, 2_{\qb}, 3_{g}, 4_{g}, 5_{\gamma}, 6_{\bar{\ell}}, 7_{\ell}\right)
+ \aa^{(0)}_{\ell}\left(1_q, 2_{\qb}, 4_{g}, 3_g, 5_{\gamma}, 6_{\bar{\ell}}, 7_{\ell}\right) \;, \nonumber  \\*
&& \aa^{(0)}_{\ell \ell}\left(1_q, 2_{\qb}, 3_{g}, 4_{\gamma}, 5_{\gamma}, 6_{\bar{\ell}}, 7_{\ell}\right) =
 \aa^{(0)}_{q\ell}\left(7_{\ell}, 6_{\bar{\ell}}, 5_{g},3_{\gamma}, 4_{\gamma}, 2_{\qb}, 1_q  \right) \;,
\label{eq:zaajamp1}
\end{eqnarray}
where $\aa^{(0)}_{q}\left(1_q,2_{\qb},3_{g},4_{g},5_{\gamma},6_{\bar{\ell}},7_{\ell}\right)$ and
$\aa^{(0)}_{\ell}\left(1_q,2_{\qb},3_{g},4_{g},5_{\gamma},6_{\bar{\ell}},7_{\ell}\right)$ 
are given in Eqs.~(\ref{eq:zajjamp1}) and~(\ref{eq:zajjamp2})~-~(\ref{eq:zajjamp3}) respectively.

\bibliography{main}
\bibliographystyle{JHEP}

\end{document}